\documentclass[acmsmall]{acmart}
\AtBeginDocument{%
  }

\usepackage[utf8]{inputenc}

\setcopyright{cc}
\setcctype{by}
\acmDOI{10.1145/3839541}
\acmYear{2026}
\acmJournal{PACMPL}
\acmVolume{10}
\acmNumber{OOPSLA2}
\acmArticle{409}
\acmMonth{10}
\acmSubmissionID{oopslab26main-p1986-p}
\received{2026-03-23}
\received[accepted]{2026-08-06}

\usepackage{xcolor}
\usepackage{listings}
\usepackage{mathtools}
\usepackage{multicol}
\usepackage{stmaryrd} 
\usepackage{bcprules}
\IfFileExists{latexml.sty}
  {\usepackage{latexml}}
  {\newif\iflatexml\latexmlfalse}
\iflatexml
  \renewcommand{\layoutruleverbose}[2]{%
    \begin{center}$\displaystyle #2$\quad #1\end{center}}
  \renewcommand{\layoutrulenolabel}[1]{%
    \begin{center}$\displaystyle #1$\end{center}}
  \newcommand{\codesize}{}
  \lxRequireResource[type=text/css]{html-fixes.css}
\else
  \newcommand{\codesize}{\small}
\fi
\usepackage{placeins}
\usepackage{enumitem}
\usepackage{booktabs}
\usepackage{subcaption}
\usepackage{pgfplots}
\pgfplotsset{compat=1.18}
\setlist[itemize]{leftmargin=1.5em, itemsep=0.3em}
\newcommand{\bind}{\mathchoice{\mathbin{\!\mapsto\!}}{\mathbin{\!\mapsto\!}}{\mapsto}{\mapsto}}
\begin{document}

\title{First-Class Refinement Types for Scala}

\author{Matt Bovel}
\correspondingauthor
\orcid{0009-0005-5132-0279}
\affiliation{%
  \institution{EPFL}
  \city{Lausanne}
  \country{Switzerland}
}

\author{Viktor Kunčak}
\orcid{0000-0001-7044-9522}
\affiliation{%
  \institution{EPFL}
  \city{Lausanne}
  \country{Switzerland}
}

\author{Martin Odersky}
\orcid{0009-0005-3923-8993}
\affiliation{%
  \institution{EPFL}
  \city{Lausanne}
  \country{Switzerland}
}

\begin{abstract}
Refinement types---types qualified with logical predicates---have proven effective for lightweight verification in languages like Liquid Haskell, F*, and Dafny. However, in these systems refinements are either written in a separate specification language or treated as second-class annotations, disconnected from the host language's type system. This disconnect creates usability barriers: programmers must maintain two mental models, and refinements cannot interact with features like type inference, subtyping, or overloading.

We present the design of first-class refinement types for Scala~3, where refinements are ordinary types that participate in subtyping, inference, and pattern matching alongside existing language features. We prove type soundness of a core, pure calculus mechanized in Rocq, combining dependent function types, bounded polymorphism, positive equi-recursive types, union and intersection types, and refinement types, using a fuel-bounded definitional interpreter and semantic typing. A distinctive design choice is our partial-correctness semantics: predicates are arbitrary terms that may diverge, and type soundness requires no termination assumptions. Finally, we implement our design as a prototype extension of the Scala~3 compiler with a lightweight e-graph-based solver for predicate entailment.
\end{abstract}



\maketitle

\section{Introduction}

A \emph{refinement type} restricts a base type to values satisfying a logical predicate. For example, the type \(\{x\!:\!\mathit{Int} \mid x > 0\}\) denotes the positive integers. Logical refinements as subsets of values have a long history~\cite{jhala2021}, going back at least to Cartwright~\cite{cartwright1976}. Freeman and Pfenning~\cite{freeman1991} coined the term \emph{refinement types} for a more restricted mechanism distinguishing subtypes of algebraic datatypes by constructor structure. Flanagan~\cite{flanagan2006} returned to general predicate refinements, checking them statically when possible and dynamically otherwise (\emph{hybrid type checking}). Rondon et al.~\cite{rondon2008} then made predicate refinements more practical with \emph{Liquid Types}, which automatically infer refinements via predicate abstraction and discharge proof obligations with SMT solvers. Systems built on these ideas such as Liquid Haskell~\cite{vazou2014}, F*~\cite{swamy2016}, and Dafny~\cite{leino2010} have since verified properties ranging from safe array indexing~\cite{rondon2008} to the functional correctness of production cryptographic libraries~\cite{zinzindohoue2017}.

The appeal of refinement types is that refinements are just \emph{types}: they are checked statically, composed through ordinary type constructors, and propagated by standard typing rules. For instance, retrieving an element from a list of non-empty strings yields a non-empty string by parametric polymorphism alone. More fundamentally, refinement types can be seen as~\cite{jhala2019why} a generalization of Floyd-Hoare logic in which invariants are attached to \emph{types} rather than \emph{program points}, delegating abstraction and instantiation to the type system using subtyping and type inference. In practice, programmers can specify and verify properties using the type system they already know, with the same error messages, IDE support, and mental model.

However, these benefits are often diminished because existing refinement type systems process refinements in a \emph{separate phase}, disconnected from the host language's type checker. In Liquid Haskell, for instance, refinements are written inside special comments and verified by an external tool. Note how the type of \lstinline!x! is declared twice in the following snippet: once as \lstinline!Int! for the Haskell type checker, and once as \lstinline!{v:Int | v mod 2 == 0}! for Liquid Haskell:
\begin{lstlisting}[language=Haskell,escapechar=!]
!\color{gray}\ttfamily\{-@ x :: \{v:Int | v mod 2 == 0 \} @-\}!
let x = 42 :: Int in ...
\end{lstlisting}
As a result, refinement types are not ``just types'': programmers must maintain two mental models, one for the host type system and one for the refinement layer, with separate error messages, separate IDE support, and limited ability to debug failures across the boundary. A recent usability study by Gamboa et al.~\cite{gamboa2025} confirms this: one participant observed that ``comments are usually seen as just optional information in the code and not something that is directly used by the compiler,'' and another remarked, ``It's sort of like you're doing two things at once because you're implementing in Haskell. But you're also talking to GHC, but you're also talking to LiquidHaskell.''

The consequences of this separation run deeper than usability. Because the refinement checker operates after or alongside the host type checker, it must duplicate much of the host's type inference machinery, an engineering burden that also limits expressiveness. Schmid and Kun\v{c}ak~\cite{schmid2016} implemented qualified types for an earlier version of the Scala compiler (Dotty) with a refinement checker ``largely independent of Scala's own type checker.'' This architecture meant refinement types could not interact with Scala's type inference (they could not be inferred as type arguments in generic code) and required a separate qualifier inference algorithm that proved difficult to scale.

Userland Scala libraries such as Iron~\cite{iron} and Refined~\cite{refined} take a different trade-off: they encode refinement-like checks using opaque types and implicit evidence, benefiting from Scala's own type inference and tooling, but at the cost of expressiveness; refinement proofs are limited to what Scala's implicit resolution can discharge, ruling out dedicated efficient decision procedures for equality, arithmetic, ordering, or other domain-specific theories.

\subsection{First-class Refinement Types}

We show that refinement types can be first-class types in a mainstream programming language. Rather than adding a separate refinement layer, we extend the Scala~3 type checker itself so that refinement types are ordinary Scala types, checked during the initial type-checking phase and flowing naturally through subtyping and type inference. The Liquid Haskell example above becomes:
\begin{lstlisting}[language=Scala]
val x: (Int with x % 2 == 0) = 42
\end{lstlisting}
Here, \lstinline!Int with x % 2 == 0! is a regular Scala type: it participates in subtyping, generics, pattern matching, type bounds, overloading, and implicits, eliminating the two-mental-models problem. There is one type system, one set of error messages, and one inference algorithm.

Integrating refinement types into the host language's type system is not entirely new: Dafny~\cite{leino2010} and F*~\cite{swamy2016} already treat refinement types as first-class citizens of their type checkers, and use subtyping to relate a refinement type to its base type. However, these languages were designed from the ground up around verification. To our knowledge, ours is the first integration of refinement types into a pre-existing language where subtyping is central and pervasive, relied upon by classes, traits, mixin composition, path-dependent types, bounded polymorphism, overloading, and implicit resolution, while preserving the type system's existing behavior.

As a concrete example, consider Scala's bounded type polymorphism. In the snippet below, the ordering is defined for \lstinline!Int! only, yet \lstinline!maximum! applied to even numbers should return \lstinline!Even!; this challenge motivated \emph{abstract refinements}~\cite{vazou2013}. In our system, bounded polymorphism and subtyping suffice: \lstinline!List(2, 4, 6)! is typed as \lstinline!List[2 | 4 | 6]!, type inference instantiates \lstinline!T! to \lstinline!Int!, whose \lstinline!Ordering! instance performs the comparisons, and \lstinline!U! to \lstinline!2 | 4 | 6!, and the result checks against \lstinline!Even! since each literal satisfies the predicate:

\begin{figure}[H]
\begin{lstlisting}[language=Scala]
def maximum[T: Ordering, U <: T](xs: List[U]): U = xs.reduce(max)
type Even = {v: Int with v % 2 == 0}
def example1: Even = maximum(List(2, 4, 6))
\end{lstlisting}
\vspace{-1em}
\caption{A refinement type interacting with bounded type polymorphism.}
\label{fig:example1}
\end{figure}

As another example, consider overload resolution, a feature characteristic of object-oriented languages. Below, we define two overloads of \lstinline!min!: one for sorted lists and one for arbitrary lists. Because refinement types participate in overload resolution, the compiler dispatches \lstinline!min(l)! to the more specific overload when \lstinline!l! is known to be sorted:

\begin{figure}[H]
\begin{lstlisting}[language=Scala]
def min(l: List[Int] with l.isSorted): Int = l.head // O(1)
def min(l: List[Int]): Int = l.min // O(n)
def example2(l: List[Int] with l.isSorted): Int = min(l) // calls the first overload
\end{lstlisting}
\vspace{-1em}
\caption{Overload resolution with refinement types.}
\label{fig:example2}
\end{figure}

\subsection{Contributions}

These examples illustrate a broader point: when refinement types are first-class, they compose freely with existing language features rather than requiring bespoke extensions for each interaction. We demonstrate this idea end-to-end, through the following contributions:
\begin{itemize}
  \item We present the \textbf{design} of first-class refinement types in Scala~3: their syntax, semantics, and interaction with Scala's existing type system, including subtyping, type inference, bounded polymorphism, overloading, and implicits (\S\ref{sec:design}). A distinctive design choice is a partial-correctness semantics for predicates: predicates are ordinary, possibly diverging pure terms whose meaning is defined operationally rather than by translation into a classical logic. A refinement holds if its predicate evaluates to $\texttt{true}$ whenever it terminates (\S\ref{sec:termination}).
  \item We prove \textbf{type soundness} of a core, pure calculus capturing the essential features of our design, mechanized in Rocq~\cite{rocq} (\S\ref{sec:formalization}). The operational semantics is given by a fuel-bounded definitional interpreter, and soundness is established via semantic typing, where types are interpreted as predicates on values. To our knowledge, this is the first mechanized soundness proof that combines refinement types with union and intersection types, bounded polymorphism (with lower and upper bounds), and positive equi-recursive types, a combination needed to model Scala's type system but not covered by prior mechanizations~\cite{hamza2019, borkowski2022, sun2024}.
  \item We describe a \textbf{prototype implementation} in the Scala~3 compiler (Dotty), showing that refinement types can be added to an industrial-strength compiler with modest changes to its architecture (\S\ref{sec:implementation}). We also present a lightweight e-graph-based solver for predicate entailment and evaluate our system on a suite of benchmarks (\S\ref{sec:evaluation}).
\end{itemize}

\section{Design}\label{sec:design}

In this section, we present the syntax (\S\ref{sec:syntax}), define the predicate language and its purity restrictions (\S\ref{sec:purity}), discuss the partial-correctness semantics (\S\ref{sec:termination}), explain how expressions are given refinement types through equality facts and selfification (\S\ref{sec:inference}), and describe run-time checks (\S\ref{sec:runtime}).

\subsection{Syntax}\label{sec:syntax}

We illustrate the syntax with the classic length-indexed vectors (Figure~\ref{fig:vec}).
\begin{figure}[H]
\begin{lstlisting}[language=Scala,morekeywords={extension}]
type Vec[T]
object Vec:
  def fill[T](n: Int, v: T): {r: Vec[T] with r.len == n} = ???
extension [T](a: Vec[T])
  def len: Int = ???
  def concat(b: Vec[T]): {r: Vec[T] with r.len == a.len + b.len} = ???
  def zip[S](b: Vec[S] with b.len == a.len): {r: Vec[(T, S)] with r.len == a.len} = ???
def example3(n: Int, m: Int): {r: Vec[(String, Int)] with r.len == m + n} =
  val v1 = Vec.fill(n, 0)
  val v2 = Vec.fill(m, 1)
  val v3 = v1.concat(v2)
  val mPlusN = m + n // inferred as Int
  Vec.fill(mPlusN, "").zip(v3)
\end{lstlisting}
\vspace{-1em}
\caption{Length-indexed vectors using refinement types.}
\label{fig:vec}
\end{figure}

A refinement type has two syntactic forms: a \emph{long form} with an explicit binder, and a \emph{short form} that reuses a name from context.

\paragraph{Long form.} The long form \lstinline!{x: T with p(x)}! introduces an explicit binder~\lstinline!x! for the refined value. This is necessary when no name is already available, such as the return type of \lstinline!fill! in Figure~\ref{fig:vec}. The curly braces and explicit binder mirror set-builder notation and are visually consistent with other brace-delimited constructs in Scala.

\paragraph{Short form.} When the refined value already has a name, as in a \lstinline!val! or parameter declaration (as for the \lstinline!b! parameter of \lstinline!zip! in Figure~\ref{fig:vec}), the binder can be omitted. The short form \lstinline!T with p! reuses the enclosing binding's name and desugars to the long form:
\begin{lstlisting}[language=Scala]
val x: (Int with x % 2 == 0) = 42
// desugars to:
val x: {v: Int with v % 2 == 0} = 42
\end{lstlisting}

\paragraph{Grammar.} Refinement types extend Scala's type grammar~\cite[\S13]{scala3spec} with two productions:
\begin{lstlisting}[morecomment={[l]{--}},commentstyle=\color{gray}\itshape]
Type           ::=  ...
                 |  InfixType `with' PostfixExpr         -- short form
SimpleType     ::=  ...
                 |  `{' id `:' InfixType `with' Expr `}' -- long form
\end{lstlisting}

Notably, predicates reuse Scala's existing expression syntax: they are syntactically ordinary expressions, though subject to semantic restrictions (\S\ref{sec:predicates}). The base type is an \lstinline!InfixType!, which excludes complex type forms such as match types without parentheses.


\subsection{Purity}\label{sec:predicates}\label{sec:purity}

Predicates are restricted to a pure fragment of Scala's expression language: constants, stable identifiers (references to \lstinline!val! bindings and parameters), field selections over \lstinline!val! fields, term and type applications, constructors of case classes without initializers, boolean connectives (\lstinline!&&!, \lstinline!||!, \lstinline!!!), comparison operators, and arithmetic. This is the subset of Scala formalized in~\S\ref{sec:formalization}; the implementation checks that predicates stay within this fragment. Importantly, predicates may refer to free variables from the enclosing scope, enabling dependent refinements such as the \lstinline!b.len == a.len! predicate in the \lstinline!zip! signature of Figure~\ref{fig:vec}.

Predicates must be \emph{pure}: \emph{deterministic and side-effect-free}\footnote{We conjecture that determinism alone is sufficient for soundness: benign side effects (e.g., logging) that do not affect the predicate's return value would be harmless. We leave formalizing this conjecture to future work.}. For instance, a mutable variable cannot appear in a predicate, and equality in predicates is structural, not referential:
\begin{lstlisting}[language=Scala]
var x = 3
val y: (Int with y == x) = 3 // rejected because x is mutable
class Box(val value: Int)
val b: (Box with b == Box(3)) = Box(3) // rejected because Box has reference equality
\end{lstlisting}
We do not currently enforce purity of called functions: our solver assumes all functions are pure, so using an impure function in a predicate is a silent logical error. This unsafe default is a deliberate choice to simplify gradual adoption without imposing a large upfront annotation burden.

A concrete path toward enforcing purity is offered by \emph{capture checking}~\cite{xu2025}, a new, experimental extension of the Scala~3 compiler.\footnote{The extensions are described in the Scala~3 documentation: \url{https://nightly.scala-lang.org/docs/reference/experimental/capture-checking/}.} It tracks \emph{capabilities}: references that confer the authority to perform effects, such as file-system or network access. The types of values and methods record which capabilities they capture or use; a \emph{pure} function is one that captures nothing. Three further extensions---\emph{Stateful Capabilities}, \emph{separation checking}~\cite{xu2026}, and \emph{safe mode}~\cite{odersky2026}---together extend this accounting to mutable state: reads and writes are tracked as capability uses, written state is exclusive and un-aliased, and escape hatches such as unchecked casts and reflection are closed, so that no effect or mutation goes untracked.

With these checks in place, purity checking of predicates reduces to a simple condition: reject any predicate that references a capability, calls a method that uses one, or reads a mutable variable. Weaker checks are also conceivable: tracking the \emph{read set} of each predicate (the read-only capabilities it uses) and invalidating established refinement facts whenever an element of that set is written to. Separation checking would make such invalidation sound, as all writes are explicit and un-aliased. Similar ideas underpin refinement type systems for Rust, where ownership likewise guarantees exclusive access to written state (\S\ref{sec:related-work}). We leave these integrations to future work.

\subsection{Non-termination and Partial Correctness}\label{sec:termination}

Our system uses a \emph{partial-correctness} semantics: a term~\lstinline!e! has type \lstinline!{x: T with p(x)}! if, whenever \lstinline!e! terminates with a value~\lstinline!v!, the predicate \lstinline!p(v)! holds. No termination guarantee is required for typing. Formally, the term interpretation $\mathcal{E}\llbracket A \rrbracket$ (\S\ref{sec:interpretation}) asserts that \emph{if} a term evaluates to a value, that value satisfies the type's interpretation. This is a key difference from prior systems, designed around termination: System~FR~\cite{hamza2019} enforces total correctness and requires termination proofs for all recursive functions, Liquid Haskell requires termination of functions reflected into refinements~\cite{vazou2014}, and F*~\cite{swamy2016} and Dafny~\cite{leino2010} require specification-level computation to terminate.

Since predicates are ordinary pure terms (\S\ref{sec:purity}), they may themselves diverge. The termination requirements of the systems above stem from their reading of predicates as formulas in a classical logic, typically that of an SMT solver. We instead define their meaning operationally: $\{x : A \mid p\}$ contains those values of~$A$ for which $p$ evaluates to $\texttt{true}$ \emph{whenever it terminates}. This mirrors the semantics of run-time assertions: evaluating $p$ as a dynamic check---as refinement patterns and \lstinline!.runtimeChecked! do in practice (\S\ref{sec:runtime})---signals a violation only if $p$ evaluates to $\texttt{false}$, so a diverging check never fails. Type soundness thus requires no termination assumptions; termination resurfaces only when justifying individual entailment rules. Crucially, a diverging predicate is vacuously satisfied: it behaves like $\texttt{true}$, and can never be used to prove $\texttt{false}$.

Termination considerations are confined to a single judgment: the \emph{semantic implication} judgment $\Gamma \vDash p_1 \Rightarrow p_2$ (\S\ref{sec:interpretation}), which the subtyping rule \textsc{S-Refine} (\S\ref{sec:subtyping}) uses to compare the predicates of two refinement types. It reads both sides partially: in every well-formed environment, if $p_1$ evaluates to $\texttt{true}$ whenever it terminates, then so does $p_2$. Thus, no type checking rule needs to refer to termination; it is only needed to justify individual entailment rules, and only some of them. Here and below, a term \emph{terminates} in a given environment if its evaluation produces a value. For example, dropping a trivially true conjunct needs no termination:\footnote{The lemmas in this subsection are proved in our Rocq mechanization, in the file \texttt{ExamplesPartialSoundness.v}.}

\begin{lemma}[\textsc{And-True}]\label{lem:and-true}
  $\forall \Gamma, t. \;\Gamma \vDash (t \mathbin{\texttt{\&\&}} \texttt{true}) \Rightarrow t$.
\end{lemma}

\noindent Other rules genuinely do:

\begin{lemma}[\textsc{And-False}]\label{lem:and-false}
  $\forall \Gamma, t. \; t \text{ terminates} \implies \Gamma \vDash (t \mathbin{\texttt{\&\&}} \texttt{false}) \Rightarrow \texttt{false}$.
\end{lemma}

\noindent The termination hypothesis is necessary: writing ``$\textsf{diverge}$'' for a closed term that never terminates, we have $\neg\,(\varnothing \vDash (\textsf{diverge} \mathbin{\texttt{\&\&}} \texttt{false}) \Rightarrow \texttt{false})$, since the left-hand side diverges and is vacuously satisfied while $\texttt{false}$ is not. The fact that a ``morally $\texttt{false}$'' predicate can be vacuously satisfied may look alarming, but it does not impact soundness: a diverging predicate is indistinguishable from $\texttt{true}$, not from $\texttt{false}$, and $\texttt{true}$ proves nothing. Writing ${<:>}$ for mutual subtyping, we have:

\begin{lemma}[Diverging predicates are vacuous]\label{lem:diverge-vacuous}
  \leavevmode
  \begin{enumerate}[nosep]
    \item $\forall \Gamma, A. \; \Gamma \vDash \{x : A \mid \textsf{diverge}\} \mathrel{<:>} \{x : A \mid \texttt{true}\} \mathrel{<:>} A$.
    \item $\forall \Gamma, t, A. \; t \text{ terminates} \implies \neg\,(\Gamma \vDash t : \{x : A \mid \texttt{false}\})$.
    \item $\neg\,(\varnothing \vDash \{x : \texttt{Unit} \mid \textsf{diverge}\} <: \{x : \texttt{Unit} \mid \texttt{false}\})$.
  \end{enumerate}
\end{lemma}

\noindent A diverging \emph{term}, by contrast, can be given the type $\{x : \texttt{Unit} \mid \texttt{false}\}$: the partial-correctness interpretation is vacuous for terms that never produce a value. This is not unsound: by Lemma~\ref{lem:diverge-vacuous}(2), no computation that actually completes has this type, so any code path that would exploit such a ``proof of $\texttt{false}$'' must first force the diverging computation, and therefore never runs. Unsoundness would arise only if the computation were erased at compile time, allowing code to assume the predicate without running it.

Because the type system is sound without any termination assumption, termination can be handled permissively, as a solver concern rather than a typing one. We therefore give the implication judgment a semantic definition only, and leave developing a syntactic proof system for predicate entailment as future work. In practice, few entailment rules are termination-sensitive: within the checked predicate fragment (\S\ref{sec:purity}), every construct except function calls trivially terminates, and termination only matters for specific rules, such as the simplification of $t \mathbin{\texttt{\&\&}} \texttt{false}$ to $\texttt{false}$ above, or the use of a predicate from a called function's result type. Even with a termination checker in place, proofs would be required only for the specific functions those rules touch---never systematically for every function used in a predicate. We believe this matters for adoption: Scala is a general-purpose language, and systematically requiring termination proofs would create a significant barrier. For now, our solver (\S\ref{sec:solver}) simply assumes termination of the functions called in predicates, as it does purity---a deliberately permissive first approximation. Verifying the termination-sensitive entailment rules and adding a termination checker are future work.

\subsection{Mixed-precision Type Inference}\label{sec:inference}

The previous sections describe how refinement types are \emph{written} and what they \emph{mean}. This section addresses how expressions actually \emph{receive} refinement types during type checking. A key design goal is backward compatibility: enabling refinement types should not change the inferred types of existing code.

\paragraph{Why not always infer precise types?}
A natural approach would be to infer refinement types everywhere. For example, systematically give \lstinline!val x = 1 + 2! its most precise type \lstinline!{v: Int with v == 1 + 2}! rather than \lstinline!Int!. This is impractical for three reasons.

First, \emph{backward compatibility}. In Scala, types are not only descriptive---they drive program elaboration. Implicit resolution and overload resolution both depend on the inferred types of arguments. A more precise type can change which implicit instances are found or introduce ambiguities between overloaded methods, breaking code that previously compiled.

Second, \emph{performance}. Carrying precise types for every intermediate expression increases the size of types, leading to higher memory usage and slower type comparisons.

Third, \emph{usability}. Overly detailed inferred types---such as deeply nested unions of singleton types---are harder to read and understand than simple approximations like \lstinline!Int!.

\paragraph{Our solution: equality facts and selfification.}
Instead of always inferring precise types, we keep Scala's existing widening behavior and recover precision on demand through two mechanisms, both standard in refinement type systems~\cite{hamza2019,borkowski2022,sun2024}.

The first mechanism is the \emph{equality fact context}. The \textsc{T-Let} typing rule (Figure~\ref{fig:typing}) records an equality fact $x \sim e_1$ between each binding $x$ and its definition $e_1$, where $\sim$ denotes that two terms evaluate to the same value:
\[
\frac{\Gamma \vdash e_1 : T_1 \qquad
  \Gamma, x : T_1,\; x \sim e_1 \vdash e_2 : T_2}
     {\Gamma \vdash \texttt{let}\; x = e_1\; \texttt{in}\; e_2 : \text{avoid}(T_2, x)}
\qquad\text{(\textsc{T-Let})}
\]
The binding \lstinline!val mPlusN = m + n! in Figure~\ref{fig:vec} is inferred as \lstinline!Int!, but the equality $\mathtt{mPlusN} \sim \mathtt{m + n}$ is recorded in the context. When \lstinline!mPlusN! is later used where a refined type is expected, the system can substitute \lstinline!m + n! for \lstinline!mPlusN! in predicates and verify the obligation. $\text{avoid}$ is described in \S\ref{sec:typing}.

The second mechanism is \emph{selfification} (\textsc{T-Self} in Figure~\ref{fig:typing}). When checking an expression~\lstinline!e! of type~\lstinline!T! against a refinement type, the compiler assigns it the self-referencing type \lstinline!{x: T with x == e}!, provided \lstinline!e! is a valid predicate term. This lifts expressions into types on demand without changing inference for unannotated code.

\subsection{Run-time Checks}\label{sec:runtime}

Refinement predicates are checked statically by default, but when the compiler cannot verify a predicate, two run-time mechanisms are available.

\paragraph{Pattern matching.} A refinement type can be used as a pattern, branching on whether a value satisfies the predicate at run time~\cite{quentin}:
\begin{lstlisting}[language=Scala]
type ID = {s: String with s.matches(idRegex)}

"a2e7-e89b" match
  case id: ID => ... // predicate holds, id has type ID
  case _      => ... // predicate does not hold
\end{lstlisting}
This integrates naturally with Scala's existing pattern-matching infrastructure. Since Scala compiles to the JVM, which erases type parameters at run time, matching against parameterized refinement types such as \lstinline!List[ID]! is not directly supported, though straightforward workarounds exist (e.g., filtering a collection with the dedicated method).

\paragraph{Checked casts.} When the programmer expects a predicate to hold but the solver cannot prove it statically, \lstinline!.runtimeChecked!~\cite{valentin} performs a dynamic check and throws an exception on failure:
\begin{lstlisting}[language=Scala]
val id: ID = "a2e7-e89b".runtimeChecked
// desugars to:
val id: ID =
  if ("a2e7-e89b".matches(idRegex)) "a2e7-e89b".asInstanceOf[ID]
  else throw new IllegalArgumentException()
\end{lstlisting}
As with other Scala types, \lstinline!.asInstanceOf! can also be used directly to skip the check entirely, at the programmer's risk.
Note that pattern matching on refinement types and \lstinline!.runtimeChecked! only works for first-order predicates that can be evaluated directly; higher-order values such as functions cannot be checked eagerly, sidestepping the blame assignment problem~\cite{keil2015} for now.

\section{Metatheory}\label{sec:formalization}

We prove type soundness for a core calculus that captures the essential features of our design. The calculus extends System~$F_{<:>}$ (System~$F_{<:}$~\cite{cardelli1991} extended with lower bounds on type variables~\cite{amin2017}) with booleans, ints, dependent function types, dependent pairs (sigma types), sum types, union and intersection types, refinement types, loops, and positive equi-recursive types. The mechanization is carried out in Rocq~\cite{rocq} using a fuel-bounded definitional interpreter~\cite{amin2017} for the operational semantics, and semantic typing~\cite{timany2024} to prove type safety.

To illustrate the scope of the calculus, Figure~\ref{fig:collect} shows a Scala function that collects elements satisfying a predicate into a refined list, alongside its encoding in the core calculus. The function exercises bounded polymorphism ($S >: T$), higher-order predicates ($p$), refinement types ($\mathit{List}[\{v : T \mid p(v)\}]$), equi-recursive types (the list itself, encoded as $\mu X.\, \texttt{Unit} + (T \times X)$), and tail-recursive iteration (encoded as $\textsf{loop}$). The mechanization includes this encoding. Running it through the interpreter for example verifies that applying it to $[3, -1, 4, -1, 5]$ with $\lambda x.\, x > 0$ yields $[3, 4, 5]$.

\begin{figure*}[t]
\small
\begin{minipage}[t]{0.49\textwidth}
\noindent\textbf{Scala source}

\vspace{0.5em}
\begin{lstlisting}[language=Scala,basicstyle=\ttfamily\codesize,morekeywords={@tailrec,then}]
enum List[T]:
  case Nil()
  case Cons(head: T, tail: List[T])

@tailrec def collect[T, S >: T](
  xs: List[T],
  p: S => Boolean,
  acc: List[{v: T with p(v)}]
): List[{v: T with p(v)}] =
  xs match
    case Nil() => acc
    case Cons(x, xs1) =>
      if p(x)
      then collect(xs1, p, Cons(x, acc))
      else collect(xs1, p, acc)
\end{lstlisting}

\end{minipage}%
\hfill%
\begin{minipage}[t]{0.49\textwidth}
\noindent\textbf{Core calculus encoding}

\vspace{1em}

$\mathit{List}[T] \triangleq \mu X.\, \texttt{Unit} + (T, X)$

\vspace{2.5em}

$\Lambda (T :> \bot <: \top).\; \Lambda (S :> T <: \top).$

\quad$\lambda \mathit{xs}{:}\mathit{List}[T].$

\quad$\lambda p{:}(S \to \texttt{Bool}).$

\quad$\lambda \mathit{acc}{:}\mathit{List}[\{v : T \mid p(v)\}].$

\qquad$\textsf{loop}((\mathit{xs},\, \mathit{acc}))\; \mathit{state}.\;$

\qquad\quad$\textsf{match}\; \mathit{state} \;\textsf{with}\; (\mathit{rem},\, \mathit{cacc}) \Rightarrow$

\qquad\qquad$\textsf{match}\; \mathit{rem} \;\textsf{with}$

\qquad\qquad\quad$\textsf{inl}(\_) \Rightarrow \textsf{inr}(\mathit{cacc})$

\qquad\qquad$\mid\; \textsf{inr}(c) \Rightarrow$

\qquad\qquad\quad$\textsf{match}\; c \;\textsf{with}\; (\mathit{hd},\, \mathit{tl}) \Rightarrow$

\qquad\qquad\qquad$\textsf{if}\; p\;\mathit{hd}$

\qquad\qquad\qquad$\textsf{then}\; \textsf{inl}((\mathit{tl},\, \textsf{inr}((\mathit{hd},\, \mathit{cacc}))))$

\qquad\qquad\qquad$\textsf{else}\; \textsf{inl}((\mathit{tl},\, \mathit{cacc}))$

\end{minipage}

\vspace{0.5em}
\caption{A \texttt{collect} function in Scala (left) and its encoding in the core calculus (right). The tail-recursive loop becomes a $\textsf{loop}$ combinator whose body returns $\textsf{inl}(\mathit{state'})$ to continue or $\textsf{inr}(\mathit{result})$ to break. Lists are equi-recursive: $\textsf{inl}(\texttt{unit})$ represents the empty list and $\textsf{inr}((\mathit{hd}, \mathit{tl}))$ a cons cell.}\label{fig:collect}
\end{figure*}

\subsection{Language}\label{sec:language}

The abstract syntax is given in Figure~\ref{fig:syntax}. Types and terms are mutually inductive, since refinement types $\{x : A \mid p\}$ embed a term~$p$ as a predicate.

\begin{figure*}[t]
\small
\begin{flushleft}
\begin{multicols}{2}\noindent\vspace{-1.5em}
\begin{flalign*}
	c \coloneqq\ & \texttt{unit} \mid \texttt{true} \mid \texttt{false} \mid z \in \mathbb{Z}  \tag*{\textbf{Constant}} \\
	%
	\iflatexml
	\mathit{op} \coloneqq\ & \texttt{==} \mid \texttt{!=} \mid \texttt{<} \mid \texttt{<=} \mid \texttt{>=} \mid \texttt{>} \mid \texttt{\&\&} \mid \texttt{||} \mid \texttt{+} \mid \texttt{-} \mid \texttt{*} \mid \texttt{/} \mid \texttt{\%}  \tag*{\textbf{Operation}} \\
	\else
	\mathit{op} \coloneqq\ & \texttt{==} \mid \texttt{!=} \mid \texttt{<} \mid \texttt{<=} \mid \texttt{>=} \mid \texttt{>}  \tag*{\textbf{Operation}} \\
	& \texttt{\&\&} \mid \texttt{||} \mid \texttt{+} \mid \texttt{-} \mid \texttt{*} \mid \texttt{/} \mid \texttt{\%}  \\
	\fi
	a, b, f, p \coloneqq\ &  \tag*{\textbf{Term}} \\
	& c  \tag*{constant} \\
	& x  \tag*{variable} \\
	& \lambda x{:}A.\, b  \tag*{abstraction} \\
	& f\; a  \tag*{application} \\
	& \Lambda (X {:>} L {<:} U).\, b  \tag*{type abstraction} \\
	& f\,[A]  \tag*{type application} \\
	& \textsf{let}\; x{:}A = a \;\textsf{in}\; b  \tag*{let} \\
	& (a_1, a_2)  \tag*{pair} \\
	\iflatexml
	& \textsf{match}\; a \;\textsf{with}\; (x, y) \Rightarrow b  \tag*{match pair} \\
	& \textsf{match}\; a \;\textsf{with}\; \textsf{inl}(x) \Rightarrow b_l \mid \textsf{inr}(y) \Rightarrow b_r  \tag*{match sum} \\
	\else
	&  \tag*{match pair} \\
	& \textsf{match}\; a \;\textsf{with}\; (x, y) \Rightarrow b \\
	&  \tag*{match sum} \\
	& \textsf{match}\; a \;\textsf{with}\; \textsf{inl}(x) \Rightarrow b_l \mid \textsf{inr}(y) \Rightarrow b_r \\
	\fi
	& \textsf{inl}[A]\, a \tag*{left injection} \\
	& \textsf{inr}[A]\, a \tag*{right injection} \\
	& a \;\mathit{op}\; b  \tag*{binary operation} \\
	& \textsf{if}\; a \;\textsf{then}\; b_1 \;\textsf{else}\; b_2  \tag*{conditional} \\
	& \textsf{loop}(a)\; x.\; b  \tag*{loop} \\[0.5em]
	\rho \coloneqq\ &  \tag*{\textbf{Environment}} \\
	& \varnothing  \tag*{empty} \\
	& \rho, x \mapsto v  \tag*{binding}
\end{flalign*}

\begin{flalign*}
	v \coloneqq\ &  \tag*{\textbf{Value}} \\
	& c  \tag*{constant} \\
	& (v_1, v_2)  \tag*{pair} \\
	& \textsf{inl}(v) \tag*{left injection} \\
	& \textsf{inr}(v) \tag*{right injection} \\
	& \langle \rho, \lambda x.\, b \rangle  \tag*{closure} \\
	& \langle \rho, \Lambda X.\, b \rangle  \tag*{type closure} \\[0.5em]
	A, B \coloneqq\ &  \tag*{\textbf{Type}} \\
	& X  \tag*{variable} \\
	& \texttt{Unit} \mid \texttt{True} \mid \texttt{False} \mid \texttt{Int32}  \tag*{base types} \\
	& \top \tag*{top} \\
	& \bot  \tag*{bottom} \\
	& \Pi x{:}A.\, B  \tag*{dependent\ function} \\
	& \forall (X {:>} L {<:} U).\, A  \tag*{bounded $\forall$} \\
	& \Sigma x{:}A.\, B  \tag*{dependent\ pair} \\
	& A + B  \tag*{sum} \\
	& \{x : A \mid p\}  \tag*{refinement} \\
	& A \lor B  \tag*{union} \\
	& A \land B  \tag*{intersection} \\
	& \mu X.\, A  \tag*{recursive} \\[0.5em]
	\Gamma \coloneqq\ &  \tag*{\textbf{Context}} \\
	& \varnothing  \tag*{empty} \\
	& \Gamma, x : A  \tag*{term binding} \\
	& \Gamma, X :> L <: U  \tag*{type var.\ bound} \\
	& \Gamma, a_1 \sim a_2  \tag*{equality fact}
\end{flalign*}
\end{multicols}
\end{flushleft}
\vspace{-2em}
\caption{Abstract syntax of the core calculus. Types and terms are mutually inductive; bindings use de Bruijn indices.}\label{fig:syntax}
\end{figure*}

\paragraph{Naming convention.}
The mechanization uses de~Bruijn indices throughout. For readability, the paper presentation uses named variables and omits explicit shifts; all rules should be read modulo $\alpha$-equivalence.

\paragraph{Terms.}
Bounded type abstractions $\Lambda (X {:>} L {<:} U).\, b$ carry both a lower bound~$L$ and an upper bound~$U$, as in $F_{<:>}$, to model Scala's bounded polymorphism.

$\textsf{loop}(a)\; x.\; b$ is a limited form of general recursion that suffices to express tail recursion: the initial state~$a$ is bound to~$x$ in the body~$b$, which returns $\textsf{inl}(v_1)$ to continue with a new state, or $\textsf{inr}(v)$ to exit with result~$v$.

Integer arithmetic uses signed 32-bit wrapping semantics ($\texttt{Int32}$) rather than mathematical integers; this matches Scala's runtime model and could in the future serve as a foundation for proving the solver's normalization rules correct. It otherwise has no impact on the formalization.

\paragraph{Values.}
As in other definitional interpreter developments~\cite{owens2016, amin2017}, values are distinct from terms: while term lambdas $\lambda x{:}A.\, b$ may have free variables, value closures $\langle \rho, \lambda x.\, b \rangle$ capture their environment~$\rho$ as a list of values. This separation reflects the big-step evaluation strategy: evaluation produces values with closed environments, not terms.

\paragraph{Types.}
The type language includes dependent function types $\Pi x{:}A.\, B$, dependent pairs $\Sigma x{:}A.\, B$, bounded universal types $\forall (X {:>} L {<:} U).\, A$, sum types, union and intersection types, and refinement types $\{x : A \mid p\}$ where the predicate~$p$ is an arbitrary term. Refinement types can be nested, as in the implementation.

The base types include singleton types $\texttt{True}$ and $\texttt{False}$ rather than a single $\texttt{Bool}$ because this enables defining the refinement type interpretation succinctly: a value satisfies $\{x : A \mid p\}$ when it belongs to $\mathcal{V}\llbracket A \rrbracket$ and $p$ evaluates to $\texttt{true}$, which is exactly membership in $\mathcal{E}\llbracket \texttt{True} \rrbracket$. The conventional $\texttt{Bool}$ is recovered as $\texttt{True} \lor \texttt{False}$.

The type $\mu X.\, A$ denotes an equi-recursive type, interpreted as the intersection of all step-indexed approximations (see \S\ref{sec:interpretation}). Unlike iso-recursive systems, no explicit fold/unfold terms are needed; instead, the equivalence $\mu X.\, A \cong A[X \mapsto \mu X.\, A]$ is established via subtyping rules (\S\ref{sec:subtyping}).

\begin{figure*}[t]
\small
\typicallabel{E-MatchPair}

\noindent\textbf{Big-step evaluation rules} \hfill \fbox{$\rho \vdash a \Downarrow v$}

\begin{multicols}{2}

\infax[E-Const]
{\rho \vdash c \Downarrow c}

\infrule[E-Var]
{x \bind v \in \rho}
{\rho \vdash x \Downarrow v}

\infax[E-Abs]
{\rho \vdash \lambda x{:}A.\, b \Downarrow \langle \rho, \lambda x.\, b \rangle}

\infax[E-TAbs]
{\rho \vdash \Lambda (X {:>} L {<:} U).\, b \Downarrow \langle \rho, \Lambda X.\, b \rangle}

\infrule[E-App]
{\rho \vdash f \Downarrow \langle \rho_f, \lambda x.\, b \rangle \\
 \rho \vdash a \Downarrow v_a \andalso \rho_f, x \bind v_a \vdash b \Downarrow v}
{\rho \vdash f\; a \Downarrow v}

\infrule[E-TApp]
{\rho \vdash f \Downarrow \langle \rho_f, \Lambda X.\, b \rangle \andalso \rho_f \vdash b \Downarrow v}
{\rho \vdash f\,[A] \Downarrow v}

\infrule[E-Let]
{\rho \vdash a \Downarrow v_a \andalso \rho, x \bind v_a \vdash b \Downarrow v}
{\rho \vdash \textsf{let}\; x{:}A = a \;\textsf{in}\; b \Downarrow v}

\infrule[E-Pair]
{\rho \vdash a_1 \Downarrow v_1 \andalso \rho \vdash a_2 \Downarrow v_2}
{\rho \vdash (a_1, a_2) \Downarrow (v_1, v_2)}

\infrule[E-BinOp]
{\rho \vdash a \Downarrow v_a \andalso \rho \vdash b \Downarrow v_b \\
 \textsf{evalop}(\mathit{op}, v_a, v_b) = r}
{\rho \vdash a \;\mathit{op}\; b \Downarrow r}

\infrule[E-IfTrue]
{\rho \vdash a \Downarrow \texttt{true} \andalso \rho \vdash b_1 \Downarrow v}
{\rho \vdash \textsf{if}\; a \;\textsf{then}\; b_1 \;\textsf{else}\; b_2 \Downarrow v}

\infrule[E-IfFalse]
{\rho \vdash a \Downarrow \texttt{false} \andalso \rho \vdash b_2 \Downarrow v}
{\rho \vdash \textsf{if}\; a \;\textsf{then}\; b_1 \;\textsf{else}\; b_2 \Downarrow v}

\infrule[E-Inl]
{\rho \vdash a \Downarrow v}
{\rho \vdash \textsf{inl}[B]\, a \Downarrow \textsf{inl}(v)}

\infrule[E-Inr]
{\rho \vdash a \Downarrow v}
{\rho \vdash \textsf{inr}[A]\, a \Downarrow \textsf{inr}(v)}

\infrule[E-LoopExit]
{\rho \vdash a \Downarrow v_0 \\
 \rho, x \bind v_0 \vdash b \Downarrow \textsf{inr}(v)}
{\rho \vdash \textsf{loop}(a)\; x.\; b \Downarrow v}

\infrule[E-LoopCont]
{\rho \vdash a \Downarrow v_0 \andalso \rho, x \bind v_0 \vdash b \Downarrow \textsf{inl}(v_1) \\
 \rho \vdash \textsf{loop}(v_1)\; x.\; b \Downarrow v}
{\rho \vdash \textsf{loop}(a)\; x.\; b \Downarrow v}

\end{multicols}

\infrule[E-MatchPair]
{\rho \vdash a \Downarrow (v_1, v_2) \andalso \rho, x \bind v_1, y \bind v_2 \vdash b \Downarrow v}
{\rho \vdash \textsf{match}\; a \;\textsf{with}\; (x, y) \Rightarrow b \Downarrow v}

\infrule[E-MatchSumL]
{\rho \vdash a \Downarrow \textsf{inl}(v) \andalso \rho, x \bind v \vdash b_l \Downarrow w}
{\rho \vdash \textsf{match}\; a \;\textsf{with}\; \textsf{inl}(x) \Rightarrow b_l \mid \textsf{inr}(y) \Rightarrow b_r \Downarrow w}

\infrule[E-MatchSumR]
{\rho \vdash a \Downarrow \textsf{inr}(v) \andalso \rho, y \bind v \vdash b_r \Downarrow w}
{\rho \vdash \textsf{match}\; a \;\textsf{with}\; \textsf{inl}(x) \Rightarrow b_l \mid \textsf{inr}(y) \Rightarrow b_r \Downarrow w}

\vspace{-1em}
\caption{Successful evaluation rules ($\rho \vdash a \Downarrow v$), derived from the definitional interpreter (Figure~\ref{fig:eval-interp}).}\label{fig:eval}
\end{figure*}

\paragraph{Contexts.}
A typing context~$\Gamma$ (Figure~\ref{fig:syntax}) contains three kinds of entries: term bindings $x : A$, type variable bounds $X :> L <: U$, and equality facts $a_1 \sim a_2$. Equality facts record that two terms evaluate to the same value. They are introduced by \textsc{T-Let}, which records that a let-bound variable equals its definition, and by \textsc{T-If}, which records that the condition equals $\textsf{true}$ or $\textsf{false}$ in the respective branches. These facts are used during subtyping to justify predicate entailment.

\subsection{Operational Semantics}\label{sec:operational-semantics}

The rules for successful evaluation $\rho \vdash a \Downarrow v$ are given in Figure~\ref{fig:eval}. Closures capture the environment at the point of creation: evaluating $\lambda x{:}A.\, b$ under~$\rho$ yields $\langle \rho, \lambda x.\, b \rangle$. Application evaluates the function and argument, then evaluates the body in the closure's environment extended with the argument. In rule \textsc{E-BinOp}, primitive operations are interpreted by the partial function $\textsf{evalop}(\mathit{op}, v_a, v_b)$ (\texttt{eval\_bin\_op} in the mechanization), which is undefined for ill-typed combinations of operands, making such applications stuck.

\paragraph{Erased semantics.}
Types are erased at runtime, so they have no operational semantics. This is reflected in the evaluation rules for type abstractions and applications, which simply ignore the type arguments. This matches Scala's runtime model, where types are erased on the JVM.

\paragraph{Definitional interpreter.}

\begin{figure}[t]
\small
\noindent\textbf{Definitional interpreter (excerpt)} \hfill \fbox{$\texttt{eval}\; n\; \rho\; a$}

\vspace{0.5em}
\begin{center}
\begin{minipage}{0.85\columnwidth}
\begin{lstlisting}[morekeywords={Fixpoint,match,with,end,Some,None,option,nat,list},deletekeywords={val},morecomment={[s]{(*}{*)}},commentstyle=\color{gray}\itshape,basicstyle=\ttfamily\codesize]
Fixpoint eval (fuel: nat) (env: list Value) (t: Term)
    : option (option Value) :=
  match fuel with
  | 0 => None (* timeout *)
  | S fuel' =>
    match t with
    | tabs _ b => Some (Some (vabs env b))
    | tapp f a =>
        match eval fuel' env f with
        | None => None
        | Some (Some (vabs envf b)) =>
            match eval fuel' env a with
            | Some (Some va) => eval fuel' (va :: envf) b
            | o => o (* propagate timeout or stuck *)
            end
        | _ => Some None (* stuck *)
        end
  ...
  end.
\end{lstlisting}
\end{minipage}
\end{center}

\vspace{0.5em}
\noindent\textbf{Termination} \hfill \fbox{$\rho \vdash a \Downarrow^? r$}

\vspace{0.3em}
\[\rho \vdash a \Downarrow^? r \;\triangleq\; \exists n.\; \texttt{eval}\; n\; \rho\; a = \texttt{Some}\; r\]

\vspace{0.5em}
\noindent\textbf{Successful evaluation} \hfill \fbox{$\rho \vdash a \Downarrow v$}

\vspace{0.3em}
\[\rho \vdash a \Downarrow v \;\triangleq\; \rho \vdash a \Downarrow^? \texttt{Some}\; v\]

\vspace{0.5em}
\caption{Fuel-bounded definitional interpreter, following Amin and Rompf~\cite{amin2017}. The interpreter returns \texttt{None} (timeout), \texttt{Some None} (stuck), or \texttt{Some (Some $v$)} (success).}\label{fig:eval-interp}
\end{figure}

In the mechanization, evaluation is implemented as a fuel-bounded definitional interpreter (Figure~\ref{fig:eval-interp}), following Amin and Rompf~\cite{amin2017}. The interpreter is a recursive Rocq function that returns a layered option type: \lstinline{None} for timeout (fuel exhausted), \lstinline{Some None} for a stuck state, or \lstinline{Some (Some v)} for successful evaluation to a value~$v$. We write $\rho \vdash a \Downarrow^? r$ when evaluation terminates with result~$\texttt{Some}\; r$ (either stuck or a value), and $\rho \vdash a \Downarrow v$ for the special case where $r = \texttt{Some}\; v$. The big-step rules in Figure~\ref{fig:eval} show only successful evaluation; the mechanization accounts for all three outcomes, as shown in Figure~\ref{fig:eval-interp}.

\paragraph{Fuel bounds depth, not steps.}\label{para:fuel}
The fuel parameter bounds the \emph{depth} of evaluation, not the total number of steps. When evaluating an application, for example, the argument, the function, and the body are all evaluated with the same, decremented fuel. This follows Amin and Rompf~\cite{amin2017}, but differs from approaches that decrement fuel at every reduction step~\cite{owens2016, paraskevopoulou2021}: for an application $f\,a$, those approaches partition the available fuel across the sub-evaluations ($\textit{fuel}_f + \textit{fuel}_a + \textit{fuel}_{body} \approx \textit{fuel}$), whereas ours passes the same decremented fuel to each ($\textit{fuel}_f = \textit{fuel}_a = \textit{fuel}_{body} = \textit{fuel} - 1$). The advantage is simplicity: depth-bounded fuel gives a natural termination measure for structural induction. The disadvantage is that it does not provide a global step count that can be correlated with recursive-type unfoldings, a technique commonly used to prove soundness of equi-recursive types via step-indexed logical relations~\cite{timany2024}. We side-step this limitation by making our value interpretation step-index-free (\S\ref{sec:interpretation}).

\subsection{Interpretation}\label{sec:interpretation}

The semantic framework is given in Figure~\ref{fig:interp}. We describe each component below.

\paragraph{Value interpretation.}
The semantic approach to type soundness represents types as predicates on values: a semantic type is a function $\textsf{Value} \to \textsf{Prop}$. The value interpretation $\mathcal{V}\llbracket A \rrbracket_{\delta}^{\rho}$ maps a syntactic type~$A$ to a semantic type, parameterized by a type variable assignment~$\delta$ (mapping type variables to semantic types) and a term variable assignment~$\rho$ (mapping term variables to values). The type variable assignment enables the interpretation of polymorphic types; the term variable assignment enables the interpretation of dependent types and refinement predicates. Using an explicit term variable assignment rather than a closing substitution is less standard, but is a natural fit for our environment-passing definitional interpreter.

Base types are interpreted as expected: $\mathcal{V}\llbracket \texttt{Unit} \rrbracket$ accepts only $\texttt{unit}$, $\mathcal{V}\llbracket \texttt{True} \rrbracket$ and $\mathcal{V}\llbracket \texttt{False} \rrbracket$ accept the corresponding boolean values, and $\mathcal{V}\llbracket \texttt{Int32} \rrbracket$ accepts any integer literal. A value satisfies a function type $\Pi x{:}A.\, B$ if it is a closure $\langle \rho_f, \lambda x.\, b \rangle$ such that for every argument~$v_a$ satisfying $\mathcal{V}\llbracket A \rrbracket$, the body~$b$ evaluated in the closure's environment~$\rho_f$ extended with~$v_a$ satisfies $\mathcal{E}\llbracket B \rrbracket$. Note that the body is evaluated in the closure's captured environment, not the caller's. A refinement type $\{x : A \mid p\}$ conjoins membership in $\mathcal{V}\llbracket A \rrbracket$ with the predicate~$p$ belonging to $\mathcal{E}\llbracket \texttt{True} \rrbracket$---i.e., evaluating to $\texttt{true}$ whenever it terminates. Union and intersection types are interpreted as set-theoretic union and intersection, respectively.

The recursive type $\mu X.\, A$ is interpreted as the intersection of all finite approximations $\forall n.\; F^n(v)$, where $F^0(v) = \top$ and $F^{n+1}(v) = \mathcal{V}\llbracket A \rrbracket_{\delta[X \mapsto F^n]}^{\rho}(v)$. Intuitively, $F^n$ captures the behavior of $\mu X.\, A$ up to $n$ unfoldings. A value inhabits the recursive type only if it satisfies all approximations. This interpretation is step-index-free in the sense that the final semantic type $\mathcal{V}\llbracket \mu X.\, A \rrbracket$ universally quantifies over all approximation levels~$n$, so it does not depend on any external step counter. The step indexing is confined to the internal construction of the approximation sequence.

\paragraph{Term interpretation.}
The term interpretation $\mathcal{E}\llbracket A \rrbracket_{\delta}^{\rho}(a)$ lifts the value interpretation to a predicate on terms. As described in \S\ref{sec:termination}, a term~$a$ has type~$A$ if, whenever evaluation of~$a$ terminates with a result~$r$ (i.e., $\rho \vdash a \Downarrow^? r$), then $r$ must be a value~$v$ satisfying $\mathcal{V}\llbracket A \rrbracket_{\delta}^{\rho}$. This rules out stuck states: if evaluation completes, it must produce a well-typed value. The statement is vacuously true for diverging terms, making this a partial correctness property.

\paragraph{Context well-formedness.}
The well-formedness predicate $\mathrm{wf}(\delta, \Gamma, \rho)$ requires that the assignments~$\delta$ and~$\rho$ are consistent with the context~$\Gamma$. Concretely, it requires three conditions: (1)~for every term binding $x : A$ in~$\Gamma$, the value $\rho(x)$ satisfies $\mathcal{V}\llbracket A \rrbracket_{\delta}^{\rho}$; (2)~for every type variable bound $X :> L <: U$ in~$\Gamma$, the semantic type $\delta(X)$ lies between $\mathcal{V}\llbracket L \rrbracket$ and $\mathcal{V}\llbracket U \rrbracket$; and (3)~for every equality fact $a_1 \sim a_2$ in~$\Gamma$, there exists a value~$v$ such that both $a_1$ and $a_2$ evaluate to~$v$. Condition~(3) is slightly simplified here: in the mechanization, each side of a fact carries a depth tag recording how many bindings were in scope when the fact was introduced, and the term is evaluated in the corresponding suffix of the environment. The simpler statement using the full environment can be recovered up to captured environments in closures: evaluating a term in a longer environment produces the same result, except that closures capture the environment available at their creation point. This is the same subtlety addressed by the evaluation weakening lemma (Lemma~\ref{lem:eval-weaken}).

\paragraph{Semantic typing.}
The semantic typing judgment $\Gamma \vDash a : A$ universally quantifies over all type variable assignments~$\delta$ and term variable assignments~$\rho$ that are well-formed with respect to~$\Gamma$, and requires that $\mathcal{E}\llbracket A \rrbracket_{\delta}^{\rho}(a)$ holds. In other words, in any environment consistent with the context, whenever evaluation of~$a$ terminates, the result is a value satisfying the type's interpretation.

\paragraph{Semantic subtyping.}
The semantic subtyping judgment $\Gamma \vDash A <: B$ similarly quantifies over all well-formed assignments and requires $\mathcal{V}\llbracket A \rrbracket_{\delta}^{\rho} \subseteq \mathcal{V}\llbracket B \rrbracket_{\delta}^{\rho}$, where semantic type inclusion $S_1 \subseteq S_2$ is defined as pointwise implication: $\forall v.\; S_1(v) \implies S_2(v)$. In other words, every value that satisfies~$A$ also satisfies~$B$.

\paragraph{Semantic implication.}
The semantic implication judgment $\Gamma \vDash p_1 \Rightarrow p_2$ requires that, in any well-formed environment, if $p_1$ evaluates to $\texttt{true}$ (i.e., $\mathcal{E}\llbracket \texttt{True} \rrbracket(p_1)$ holds), then so does $p_2$. This judgment is used in the refinement subtyping rule \textsc{S-Refine} to compare predicates.

\begin{figure*}[p]
\small
\noindent\textbf{Value interpretation} \hfill \fbox{$\mathcal{V}\llbracket A \rrbracket_{\delta}^{\rho}(v)$}

\vspace{0.5em}
\begin{math}
\begin{array}{r@{\;}c@{\;}l}
  \mathcal{V}\llbracket X \rrbracket_{\delta}^{\rho}(v) &\triangleq& \delta(X)(v)
  \\[0.5em]
  \mathcal{V}\llbracket \texttt{Unit} \rrbracket_{\delta}^{\rho}(v) &\triangleq& v = \texttt{unit}
  \\[0.5em]
  \mathcal{V}\llbracket \texttt{True} \rrbracket_{\delta}^{\rho}(v) &\triangleq& v = \texttt{true}
  \\[0.5em]
  \mathcal{V}\llbracket \texttt{False} \rrbracket_{\delta}^{\rho}(v) &\triangleq& v = \texttt{false}
  \\[0.5em]
  \mathcal{V}\llbracket \texttt{Int32} \rrbracket_{\delta}^{\rho}(v) &\triangleq& \exists z.\; v = z
  \\[0.5em]
  \mathcal{V}\llbracket \top \rrbracket_{\delta}^{\rho}(v) &\triangleq& \top
  \\[0.5em]
  \mathcal{V}\llbracket \bot \rrbracket_{\delta}^{\rho}(v) &\triangleq& \bot
  \\[0.5em]
  \mathcal{V}\llbracket \Pi x{:}A.\, B \rrbracket_{\delta}^{\rho}(v) &\triangleq&
    \exists \rho_f, b.\; v = \langle \rho_f, \lambda x.\, b \rangle \land \forall v_a.\; \mathcal{V}\llbracket A \rrbracket_{\delta}^{\rho}(v_a) \implies \mathcal{E}\llbracket B \rrbracket_{\delta}^{\rho_f[x \mapsto v_a]}(b)
  \\[0.5em]
  \mathcal{V}\llbracket \forall (X {:>} L {<:} U).\, B \rrbracket_{\delta}^{\rho}(v) &\triangleq&
    \exists \rho_f, b.\; v = \langle \rho_f, \Lambda X.\, b \rangle \land \forall S.\; \mathcal{V}\llbracket L \rrbracket_{\delta}^{\rho} \subseteq S \subseteq \mathcal{V}\llbracket U \rrbracket_{\delta}^{\rho} \implies \mathcal{E}\llbracket B \rrbracket_{\delta[X \mapsto S]}^{\rho_f}(b)
  \\[0.5em]
  \mathcal{V}\llbracket \Sigma x{:}A.\, B \rrbracket_{\delta}^{\rho}(v) &\triangleq&
    \exists v_1, v_2.\; v = (v_1, v_2) \land \mathcal{V}\llbracket A \rrbracket_{\delta}^{\rho}(v_1) \land \mathcal{V}\llbracket B \rrbracket_{\delta}^{\rho[x \mapsto v_1]}(v_2)
  \\[0.5em]
  \mathcal{V}\llbracket A + B \rrbracket_{\delta}^{\rho}(v) &\triangleq&
    (\exists w.\; v = \textsf{inl}(w) \land \mathcal{V}\llbracket A \rrbracket_{\delta}^{\rho}(w)) \lor (\exists w.\; v = \textsf{inr}(w) \land \mathcal{V}\llbracket B \rrbracket_{\delta}^{\rho}(w))
  \\[0.5em]
  \mathcal{V}\llbracket \{x : A \mid p\} \rrbracket_{\delta}^{\rho}(v) &\triangleq& \mathcal{V}\llbracket A \rrbracket_{\delta}^{\rho}(v) \land \mathcal{E}\llbracket \texttt{True} \rrbracket_{\delta}^{\rho[x \mapsto v]}(p)
  \\[0.5em]
  \mathcal{V}\llbracket A \lor B \rrbracket_{\delta}^{\rho}(v) &\triangleq& \mathcal{V}\llbracket A \rrbracket_{\delta}^{\rho}(v) \lor \mathcal{V}\llbracket B \rrbracket_{\delta}^{\rho}(v)
  \\[0.5em]
  \mathcal{V}\llbracket A \land B \rrbracket_{\delta}^{\rho}(v) &\triangleq& \mathcal{V}\llbracket A \rrbracket_{\delta}^{\rho}(v) \land \mathcal{V}\llbracket B \rrbracket_{\delta}^{\rho}(v)
  \\[0.5em]
  \mathcal{V}\llbracket \mu X.\, A \rrbracket_{\delta}^{\rho}(v) &\triangleq& \forall n.\; F^n(v) \quad \text{where } F^0(v) = \top,\; F^{n+1}(v) = \mathcal{V}\llbracket A \rrbracket_{\delta[X \mapsto F^n]}^{\rho}(v)
\end{array}
\end{math}

\vspace{0.7em}
\noindent\textbf{Term interpretation} \hfill \fbox{$\mathcal{E}\llbracket A \rrbracket_{\delta}^{\rho}(a)$}

\vspace{0.5em}
\begin{math}
\begin{array}{r@{\;}c@{\;}l}
  \mathcal{E}\llbracket A \rrbracket_{\delta}^{\rho}(a) &\triangleq& \forall r.\; \rho \vdash a \Downarrow^? r \implies \exists v.\; r = \texttt{Some}\; v \land \mathcal{V}\llbracket A \rrbracket_{\delta}^{\rho}(v)
\end{array}
\end{math}

\vspace{0.7em}
\noindent\textbf{Context well-formedness} \hfill \fbox{$\mathrm{wf}(\delta, \Gamma, \rho)$}

\vspace{0.5em}
\begin{math}
\begin{array}{r@{\;}c@{\;}l}
  \mathrm{wf}(\delta, \Gamma, \rho) &\triangleq& \forall (x : A) \in \Gamma.\; \mathcal{V}\llbracket A \rrbracket_{\delta}^{\rho}(\rho(x))
  \\
  && \land\; \forall (X {:>} L {<:} U) \in \Gamma.\; \mathcal{V}\llbracket L \rrbracket_{\delta}^{\rho} \subseteq \delta(X) \subseteq \mathcal{V}\llbracket U \rrbracket_{\delta}^{\rho}
  \\
  && \land\; \forall (a_1 \sim a_2) \in \Gamma.\; \exists v.\; \rho \vdash a_1 \Downarrow v \land \rho \vdash a_2 \Downarrow v
\end{array}
\end{math}

\vspace{0.7em}
\noindent\textbf{Semantic typing} \hfill \fbox{$\Gamma \vDash a : A$}

\vspace{0.5em}
\begin{math}
\begin{array}{r@{\;}c@{\;}l}
  \Gamma \vDash a : A &\triangleq& \forall \delta, \rho.\; \mathrm{wf}(\delta, \Gamma, \rho) \implies \mathcal{E}\llbracket A \rrbracket_{\delta}^{\rho}(a)
\end{array}
\end{math}

\vspace{0.7em}
\noindent\textbf{Semantic subtyping} \hfill \fbox{$\Gamma \vDash A <: B$}

\vspace{0.5em}
\begin{math}
\begin{array}{r@{\;}c@{\;}l}
  \Gamma \vDash A <: B &\triangleq& \forall \delta, \rho.\; \mathrm{wf}(\delta, \Gamma, \rho) \implies \mathcal{V}\llbracket A \rrbracket_{\delta}^{\rho} \subseteq \mathcal{V}\llbracket B \rrbracket_{\delta}^{\rho}
\end{array}
\end{math}

\vspace{0.7em}
\noindent\textbf{Semantic implication} \hfill \fbox{$\Gamma \vDash p_1 \Rightarrow p_2$}

\vspace{0.5em}
\begin{math}
\begin{array}{r@{\;}c@{\;}l}
  \Gamma \vDash p_1 \Rightarrow p_2 &\triangleq& \forall \delta, \rho.\; \mathrm{wf}(\delta, \Gamma, \rho) \implies \mathcal{E}\llbracket \texttt{True} \rrbracket_{\delta}^{\rho}(p_1) \implies \mathcal{E}\llbracket \texttt{True} \rrbracket_{\delta}^{\rho}(p_2)
\end{array}
\end{math}

\vspace{0.7em}
\noindent\textbf{Semantic type inclusion} \hfill \fbox{$S_1 \subseteq S_2$}

\vspace{0.5em}
\begin{math}
\begin{array}{r@{\;}c@{\;}l}
  S_1 \subseteq S_2 &\triangleq& \forall v.\; S_1(v) \implies S_2(v)
\end{array}
\end{math}

\vspace{0.5em}
\caption{Semantic framework. The value interpretation $\mathcal{V}\llbracket \cdot \rrbracket$ maps syntactic types to predicates on values. The term interpretation $\mathcal{E}\llbracket \cdot \rrbracket$ lifts this to a partial correctness assertion on terms. Semantic typing and subtyping quantify over all well-formed environments.}\label{fig:interp}
\end{figure*}

\subsection{Typing}\label{sec:typing}

The semantic typing judgment $\Gamma \vDash a : A$ (Figure~\ref{fig:interp}) requires that, in any well-formed environment, $\mathcal{E}\llbracket A \rrbracket$ holds for~$a$---i.e., whenever evaluation terminates, the result is a value in $\mathcal{V}\llbracket A \rrbracket$.
The typing rules in Figure~\ref{fig:typing} are proven sound with respect to this judgment. Each rule corresponds to a proven lemma. Noteworthy rules include:
\begin{itemize}
  \item \textsc{T-Let} records the equality $x \sim a$ in the context, enabling the body to use facts about the bound variable's value. The result type uses $\textsf{avoid}(B, x)$ to remove references to the bound variable, which goes out of scope.
  \item \textsc{T-Self} (selfification) refines a first-order value to a singleton type $\{x : A \mid x \mathbin{\texttt{==}} a\}$. The $\textsf{firstorder}(A)$ check ensures that~$A$ is a type for which equality comparison is defined---concretely, $\texttt{Unit}$, $\texttt{True}$, $\texttt{False}$, $\texttt{Int32}$, or a refinement of one of these. Functions, closures, and polymorphic values are excluded because they do not support runtime equality.
  \item \textsc{T-If} adds equality facts $a \sim \textsf{true}$ and $a \sim \textsf{false}$ in the respective branches, and returns a union type $B_1 \lor B_2$ of the two branches' types.
  \item \textsc{T-Loop} types a loop whose body returns $A + B$: $\textsf{inl}$ continues with a new accumulator of type~$A$, $\textsf{inr}$ exits with a result of type~$B$. This rule allows recursion without typing a self-reference, avoiding the need for step-indexed reasoning that would typically be required for a general fixpoint combinator.
\end{itemize}

\paragraph{ANF restriction.}
Several typing rules require certain subterms to be variables rather than arbitrary expressions. In \textsc{T-App}, the argument must be a variable~$y$, so that the substitution $B[x \mapsto y]$ in the result type remains a syntactic type (substituting an arbitrary term could produce a type containing a non-variable subterm, complicating the interpretation). Similarly, \textsc{T-Pair} requires the first component to be a variable. This administrative normal form (ANF) restriction does not reduce expressiveness: any argument expression can be let-bound first, with the let-bound variable used as the argument. This is exactly what the implementation does via skolems (\S\ref{sec:anf}), which hoist arguments to local bindings so that dependent result types can refer to them by name.

\paragraph{Avoidance.}
The $\textsf{avoid}(A, x)$ operation removes occurrences of a bound variable~$x$ from a type~$A$, producing a well-scoped result type. It traverses the type structurally, guided by a polarity flag that flips at contravariant positions (the domain of~$\Pi$ and the upper bound of~$\forall$). When a refinement predicate mentions~$x$, it is replaced by $\texttt{true}$ in positive polarity (yielding a supertype) or $\texttt{false}$ in negative polarity (yielding a subtype). For recursive types $\mu Y.\, B$, avoidance recurses into the body only when $\textsf{spos}(Y, B)$ holds; otherwise it falls back to $\top$ or $\bot$. The soundness of this operation is established in Lemma~\ref{lem:avoid}.

\begin{figure*}[t]
\small
\typicallabel{T-MatchPair}

\noindent\textbf{Semantic typing rules} \hfill \fbox{$\Gamma \vDash a : A$}

\begin{multicols}{2}

\infax[T-Unit]
{\Gamma \vDash \texttt{unit} : \texttt{Unit}}

\infax[T-True]
{\Gamma \vDash \texttt{true} : \texttt{True}}

\infax[T-False]
{\Gamma \vDash \texttt{false} : \texttt{False}}

\infax[T-Int]
{\Gamma \vDash z : \texttt{Int32}}

\infrule[T-Var]
{x : A \in \Gamma}
{\Gamma \vDash x : A}

\infrule[T-Abs]
{\Gamma, x : A \vDash b : B}
{\Gamma \vDash \lambda x{:}A.\, b : \Pi x{:}A.\, B}

\infrule[T-App]
{\Gamma \vDash f : \Pi x{:}A.\, B \andalso \Gamma \vDash y : A}
{\Gamma \vDash f\; y : B[x \bind y]}

\infrule[T-TAbs]
{\Gamma, X :> L <: U \vDash b : B}
{\Gamma \vDash \Lambda (X {:>} L {<:} U).\, b : \forall (X {:>} L {<:} U).\, B}

\infrule[T-TApp]
{\Gamma \vDash f : \forall (X {:>} L {<:} U).\, B \\
 \Gamma \vDash L <: A \andalso \Gamma \vDash A <: U}
{\Gamma \vDash f\,[A] : B[X \bind A]}

\infrule[T-Let]
{\Gamma \vDash a : A \andalso \Gamma, x : A, x \sim a \vDash b : B}
{\Gamma \vDash \textsf{let}\; x{:}A = a \;\textsf{in}\; b : \textsf{avoid}(B, x)}

\infrule[T-Pair]
{\Gamma \vDash y : A \andalso \Gamma \vDash a_2 : B[x \bind y]}
{\Gamma \vDash (y, a_2) : \Sigma x{:}A.\, B}

\infrule[T-BinOp]
{\Gamma \vDash a : A \andalso \Gamma \vDash b : A \\
 \textsf{compat}(\mathit{op}, A)}
{\Gamma \vDash a \;\mathit{op}\; b : \textsf{result}(\mathit{op}, A)}

\infrule[T-If]
{\Gamma \vDash a : \texttt{True} \lor \texttt{False} \\
 \Gamma, a \sim \textsf{true} \vDash b_1 : B_1 \\
 \Gamma, a \sim \textsf{false} \vDash b_2 : B_2}
{\Gamma \vDash \textsf{if}\; a \;\textsf{then}\; b_1 \;\textsf{else}\; b_2 : B_1 \lor B_2}

\infrule[T-Inl]
{\Gamma \vDash a : A}
{\Gamma \vDash \textsf{inl}[B]\, a : A + B}

\infrule[T-Inr]
{\Gamma \vDash a : B}
{\Gamma \vDash \textsf{inr}[A]\, a : A + B}

\infrule[T-Loop]
{\Gamma \vDash a : A \andalso \Gamma, x : A \vDash b : A + B}
{\Gamma \vDash \textsf{loop}(a)\; x.\; b : B}

\infrule[T-Self]
{\Gamma \vDash a : A \andalso \textsf{firstorder}(A)}
{\Gamma \vDash a : \{x : A \mid x \mathbin{\texttt{==}} a\}}

\infrule[T-Sub]
{\Gamma \vDash a : A \andalso \Gamma \vDash A <: B}
{\Gamma \vDash a : B}

\end{multicols}

\infrule[T-MatchPair]
{\Gamma \vDash a : \Sigma x{:}A.\, B \andalso \Gamma, x : A, y : B \vDash b : C}
{\Gamma \vDash \textsf{match}\; a \;\textsf{with}\; (x, y) \Rightarrow b : \textsf{avoid}(\textsf{avoid}(C, y), x)}

\infrule[T-MatchSum]
{\Gamma \vDash a : A_1 + A_2 \andalso \Gamma, x : A_1, a \sim \textsf{inl}(x) \vDash b_1 : B_1 \andalso \Gamma, y : A_2, a \sim \textsf{inr}(y) \vDash b_2 : B_2}
{\Gamma \vDash \textsf{match}\; a \;\textsf{with}\; \textsf{inl}(x) \Rightarrow b_1 \mid \textsf{inr}(y) \Rightarrow b_2 : \textsf{avoid}(B_1, x) \lor \textsf{avoid}(B_2, y)}

\vspace{-1em}
\caption{Typing rules. All rules are proven sound with respect to the semantic typing judgment.}\label{fig:typing}
\end{figure*}

\subsection{Subtyping}\label{sec:subtyping}

The semantic subtyping judgment $\Gamma \vDash A <: B$ (Figure~\ref{fig:interp}) requires that, in any well-formed environment, $\mathcal{V}\llbracket A \rrbracket \subseteq \mathcal{V}\llbracket B \rrbracket$.
The subtyping rules in Figure~\ref{fig:subtyping} are proven sound with respect to this judgment. Function types are contravariant in their domain and covariant in their codomain (\textsc{S-Fun}); bounded universal types are covariant in the lower bound, contravariant in the upper bound, and covariant in the body (\textsc{S-Forall}). Refinement types support two rules: \textsc{S-RefineBase} drops the predicate (every refined type is a subtype of its base), and \textsc{S-Refine} compares both base types and predicates, using the semantic implication judgment $\Gamma \vDash p_1 \Rightarrow p_2$ to establish that the predicate of the subtype entails the predicate of the supertype.

The equi-recursive rules \textsc{S-Mu-Fold} and \textsc{S-Mu-Unfold} establish the isomorphism $\mu X.\, A \cong A[X \mapsto \mu X.\, A]$, requiring that $X$ appears only in strictly positive positions in~$A$, as defined by the predicate $\textsf{spos}(X, A)$ (Figure~\ref{fig:spos}). Strict positivity means $X$ never appears to the left of a function arrow or in the bounds of a universal type. For union types, $\textsf{spos}$ requires $X$ to be entirely absent; this stronger condition is needed for a distributing lemma used in the fold/unfold proofs. Nested recursive types $\mu Y.\, B$ require both $\textsf{spos}(Y, B)$ and $X \notin B$: the latter prevents the outer variable from leaking into the inner fixpoint's approximation sequence.

\begin{figure*}[t]
\small
\typicallabel{S-RefineBase}

\noindent\textbf{Semantic subtyping rules} \hfill \fbox{$\Gamma \vDash A <: B$}

\begin{multicols}{2}

\infax[S-Refl]
{\Gamma \vDash A <: A}

\infrule[S-Trans]
{\Gamma \vDash A <: B \andalso \Gamma \vDash B <: C}
{\Gamma \vDash A <: C}

\infax[S-Top]
{\Gamma \vDash A <: \top}

\infax[S-Bot]
{\Gamma \vDash \bot <: A}

\infrule[S-Fun]
{\Gamma \vDash B_1 <: A_1 \andalso \Gamma, x : A_1 \vDash A_2 <: B_2}
{\Gamma \vDash \Pi x{:}A_1.\, A_2 <: \Pi x{:}B_1.\, B_2}

\infrule[S-Sigma]
{\Gamma \vDash A_1 <: B_1 \andalso \Gamma, x : A_1 \vDash A_2 <: B_2}
{\Gamma \vDash \Sigma x{:}A_1.\, A_2 <: \Sigma x{:}B_1.\, B_2}

\infax[S-OrL]
{\Gamma \vDash A <: A \lor B}

\infax[S-OrR]
{\Gamma \vDash B <: A \lor B}

\infrule[S-Or]
{\Gamma \vDash A <: C \andalso \Gamma \vDash B <: C}
{\Gamma \vDash A \lor B <: C}

\infax[S-AndL]
{\Gamma \vDash A \land B <: A}

\infax[S-AndR]
{\Gamma \vDash A \land B <: B}

\infrule[S-And]
{\Gamma \vDash C <: A \andalso \Gamma \vDash C <: B}
{\Gamma \vDash C <: A \land B}

\infax[S-RefineBase]
{\Gamma \vDash \{x : A \mid p\} <: A}

\infrule[S-Refine]
{\Gamma \vDash A <: B \andalso \Gamma, x : A \vDash p_1 \Rightarrow p_2}
{\Gamma \vDash \{x : A \mid p_1\} <: \{x : B \mid p_2\}}

\infrule[S-TVar-Upper]
{X :> L <: U \in \Gamma}
{\Gamma \vDash X <: U}

\infrule[S-TVar-Lower]
{X :> L <: U \in \Gamma}
{\Gamma \vDash L <: X}

\infrule[S-Mu-Unfold]
{\textsf{spos}(X, A)}
{\Gamma \vDash \mu X.\, A <: A[X \bind \mu X.\, A]}

\infrule[S-Mu-Fold]
{\textsf{spos}(X, A)}
{\Gamma \vDash A[X \bind \mu X.\, A] <: \mu X.\, A}

\end{multicols}

\infrule[S-Forall]
{\Gamma \vDash L_1 <: L_2 \andalso \Gamma \vDash U_2 <: U_1 \andalso \Gamma, X :> L_2 <: U_2 \vDash A <: B}
{\Gamma \vDash \forall (X {:>} L_1 {<:} U_1).\, A <: \forall (X {:>} L_2 {<:} U_2).\, B}

\vspace{-1em}
\caption{Subtyping rules. All rules are proven sound with respect to the semantic subtyping judgment.}\label{fig:subtyping}
\end{figure*}

\begin{figure}[t]
\small
\noindent\textbf{Strict positivity} \hfill \fbox{$\textsf{spos}(X, A)$}

\vspace{0.5em}
\begin{math}
\begin{array}{r@{\;}c@{\;}l}
  \textsf{spos}(X, A) &\triangleq& \mathrm{True} \quad \text{if $A$ is a base type or type variable}
  \\[0.3em]
  \textsf{spos}(X, \Pi x{:}A.\, B) &\triangleq& X \notin A \land \textsf{spos}(X, B)
  \\[0.3em]
  \textsf{spos}(X, \forall (Y {:>} L {<:} U).\, B) &\triangleq& X \notin L \land X \notin U \land \textsf{spos}(X, B)
  \\[0.3em]
  \textsf{spos}(X, \mu Y.\, A) &\triangleq& \textsf{spos}(Y, A) \land X \notin A
  \\[0.3em]
  \textsf{spos}(X, \Sigma x{:}A.\, B) &\triangleq& \textsf{spos}(X, A) \land \textsf{spos}(X, B)
  \\[0.3em]
  \textsf{spos}(X, A + B) &\triangleq& \textsf{spos}(X, A) \land \textsf{spos}(X, B)
  \\[0.3em]
  \textsf{spos}(X, \{x : A \mid p\}) &\triangleq& \textsf{spos}(X, A)
  \\[0.3em]
  \textsf{spos}(X, A \lor B) &\triangleq& X \notin A \land X \notin B
  \\[0.3em]
  \textsf{spos}(X, A \land B) &\triangleq& \textsf{spos}(X, A) \land \textsf{spos}(X, B)
\end{array}
\end{math}

\vspace{0.5em}
\caption{Strict positivity. $\textsf{spos}(X, A)$ holds when $X$ appears only in strictly positive positions in $A$: never to the left of a function arrow, and never in the bounds of a universal type. Union types require $X$ to be entirely absent (needed for the distributing lemma).}\label{fig:spos}
\end{figure}

\subsection{Soundness and Main Lemmas}\label{sec:soundness}

Each typing rule in Figure~\ref{fig:typing} and each subtyping rule in Figure~\ref{fig:subtyping} is individually proven sound as a lemma about the semantic judgments $\Gamma \vDash a : A$ and $\Gamma \vDash A <: B$. Since these semantic judgments directly assert that well-typed terms do not get stuck and that subtyping preserves value membership, any derivation composed from these rules inherits the safety guarantee. This is the fundamental theorem of the logical relation.

\paragraph{Adequacy.}
To provide a conventional syntactic interface, we also define inductive syntactic typing and subtyping judgments that mirror the semantic rules of Figures~\ref{fig:typing} and~\ref{fig:subtyping}. The adequacy theorems show that these syntactic judgments imply their semantic counterparts. The proofs proceed by straightforward induction on the derivation, applying the corresponding semantic lemma at each constructor.

\begin{theorem}[Adequacy of typing]\label{thm:adequacy-typing}
  If\/ $\Gamma \vdash a : A$ then $\Gamma \vDash a : A$.
\end{theorem}

\begin{theorem}[Adequacy of subtyping]\label{thm:adequacy-subtyping}
  If\/ $\Gamma \vdash A <: B$ then $\Gamma \vDash A <: B$.
\end{theorem}

\noindent Note that the semantic implication judgment $\Gamma \vDash p_1 \Rightarrow p_2$ has no syntactic counterpart: we do not define inference rules for it. Instead, it remains an entirely semantic notion, used directly in the subtyping rule \textsc{S-Refine}. This reflects the design of the implementation, where predicate entailment is delegated to the solver (\S\ref{sec:solver}) rather than derived from syntactic rules. Syntactic entailment rules could nonetheless be proved sound against the semantic judgment, just as the typing and subtyping rules are; Lemmas~\ref{lem:and-true} and~\ref{lem:and-false} (\S\ref{sec:termination}) exemplify this, showing that some rules hold unconditionally while others require termination assumptions.

This separation between syntactic and semantic judgments allows us to state and use typing rules in the familiar inductive style, while the soundness guarantee flows from the semantic interpretation.

\paragraph{Weakening and substitution for the interpretation.}
The semantic typing and subtyping proofs require that the value interpretation is well-behaved under changes to the environment. The key lemmas establish that the interpretation depends only on the variables the type mentions: extending the environment with a fresh binding leaves it unchanged, and a variable can be substituted away when the environment determines its value. We write $\mathrm{FV}(A)$ for the free (term and type) variables of~$A$.\footnote{The mechanization states these lemmas in de~Bruijn form, where extending the environment corresponds to an explicit index shift on the type and the freshness side conditions are implicit.}

\begin{lemma}[Term weakening]\label{lem:interp-term-weaken}
  If\/ $x \notin \mathrm{FV}(A)$, then
  $\mathcal{V}\llbracket A \rrbracket_{\delta}^{\rho} = \mathcal{V}\llbracket A \rrbracket_{\delta}^{\rho[x \bind v]}$.
\end{lemma}

\begin{lemma}[Term substitution]\label{lem:interp-term-subst}
  If\/ $\rho(y) = v$, then
  $\mathcal{V}\llbracket A \rrbracket_{\delta}^{\rho[x \bind v]} = \mathcal{V}\llbracket A[x \bind y] \rrbracket_{\delta}^{\rho}$.
\end{lemma}

\begin{lemma}[Type weakening]\label{lem:interp-type-weaken}
  If\/ $X \notin \mathrm{FV}(A)$, then
  $\mathcal{V}\llbracket A \rrbracket_{\delta}^{\rho} = \mathcal{V}\llbracket A \rrbracket_{\delta[X \bind S]}^{\rho}$.
\end{lemma}

\begin{lemma}[Type substitution]\label{lem:interp-type-subst}
  $\mathcal{V}\llbracket B \rrbracket_{\delta[X \bind \mathcal{V}\llbracket A \rrbracket_\delta^\rho]}^{\rho} = \mathcal{V}\llbracket B[X \bind A] \rrbracket_{\delta}^{\rho}$
\end{lemma}

\noindent The mechanization proves more general versions for multi-element extensions, from which these are derived. These lemmas are used pervasively---for instance, term substitution is needed for \textsc{T-App} and \textsc{T-Let}, and type substitution for \textsc{T-TApp} and \textsc{S-Mu-Fold}/\textsc{S-Mu-Unfold}.

\paragraph{Weakening and substitution for evaluation.}
The interpretation lemmas above reduce questions about the semantic types to questions about evaluation under modified environments. These in turn require corresponding lemmas about the evaluator. The difficulty is that evaluation results are not preserved exactly: when the environment is extended, closures capture the larger environment, producing structurally different values. To handle this, we define a \emph{compatibility relation} $r_1 \approx r_2$ on results that relates first-order values to themselves and closures to closures with correspondingly extended captured environments. The evaluation weakening lemmas then take the following simulation-style form:

\begin{lemma}[Evaluation weakening]\label{lem:eval-weaken}
  Suppose $x \notin \mathrm{FV}(a)$.
  If\/ $\rho \vdash a \Downarrow^? r$, then there exists $r' \approx r$ such that
  $\rho[x \bind v] \vdash a \Downarrow^? r'$.
  Conversely, if\/ $\rho[x \bind v] \vdash a \Downarrow^? r$, then there exists $r' \approx r$ such that $\rho \vdash a \Downarrow^? r'$.
\end{lemma}

\noindent This simulation-style argument is needed because the definitional interpreter uses closures rather than substitution: evaluating a lambda under a longer environment produces a closure capturing that longer environment. Analogous lemmas handle term substitution, with a corresponding substitution compatibility relation.

\paragraph{Avoidance.}
The $\textsf{avoid}$ operation (\S\ref{sec:typing}) must preserve the subtyping relationship in the appropriate direction. The key lemma establishes this for both polarities simultaneously:

\begin{lemma}[Avoidance soundness]\label{lem:avoid}
  For all $v$:
  \begin{enumerate}[nosep]
    \item $\mathcal{V}\llbracket A \rrbracket_{\delta}^{\rho}(v) \implies \mathcal{V}\llbracket \textsf{avoid}(A, x)^{+} \rrbracket_{\delta}^{\rho}(v)$\quad (positive: supertype)
    \item $\mathcal{V}\llbracket \textsf{avoid}(A, x)^{-} \rrbracket_{\delta}^{\rho}(v) \implies \mathcal{V}\llbracket A \rrbracket_{\delta}^{\rho}(v)$\quad (negative: subtype)
  \end{enumerate}
\end{lemma}

\noindent These are proven together by induction on~$A$, flipping polarity at contravariant positions. The $\textsf{avoid}(B, x)$ appearing in the typing rules (e.g., \textsc{T-Let}) uses positive polarity, $\textsf{avoid}(B, x)^{+}$, producing a supertype of~$B$.

\paragraph{Positivity and fold/unfold.}
The fold and unfold rules for $\mu X.\, A$ require showing that the interpretation of the recursive type coincides with its one-step unfolding: $\mathcal{V}\llbracket \mu X.\, A \rrbracket = \mathcal{V}\llbracket A[X \bind \mu X.\, A] \rrbracket$. The key ingredients, adapted from Hamza et~al.~\cite{hamza2019}, are:

\begin{lemma}[Monotonicity]\label{lem:spos-mono}
  If $\textsf{spos}(X, A)$ and $S_1 \subseteq S_2$, then
  $\mathcal{V}\llbracket A \rrbracket_{\delta[X \bind S_1]} \subseteq \mathcal{V}\llbracket A \rrbracket_{\delta[X \bind S_2]}$.
\end{lemma}

\begin{lemma}[Distribution]\label{lem:spos-distrib}
  Let $S_\cap = \lambda w.\, \forall n.\, S_n(w)$.
  If $\textsf{spos}(X, A)$ and $\forall n.\; \mathcal{V}\llbracket A \rrbracket_{\delta[X \bind S_n]}(v)$, then
  $\mathcal{V}\llbracket A \rrbracket_{\delta[X \bind S_\cap]}(v)$.
\end{lemma}

\noindent Monotonicity says that the interpretation of a strictly positive type is monotone in the type variable's denotation. Distribution says that an intersection over a family of denotations can be pushed inside the interpretation. Together, these establish that the approximation sequence $F^0 \supseteq F^1 \supseteq \cdots$ converges to a fixpoint, so $\forall n.\, F^n(v)$ is equivalent to $F(\lambda w.\, \forall n.\, F^n(w))(v)$, which is exactly $\mathcal{V}\llbracket A[X \bind \mu X.\, A] \rrbracket(v)$.

\section{Implementation}\label{sec:implementation}

We implement the design of \S\ref{sec:design} as a prototype extension of Dotty, the production Scala~3 compiler. This section describes how refinement types integrate into the existing compiler architecture (\S\ref{sec:integration}), how argument hoisting and skolems recover the ANF discipline assumed by the formalization (\S\ref{sec:anf}), and the lightweight e-graph-based solver that discharges predicate entailment obligations (\S\ref{sec:solver}). We evaluate compilation overhead on a suite of benchmarks in \S\ref{sec:evaluation}.

\subsection{Integration}\label{sec:integration}

We implement refinement types in Dotty, the production Scala~3 compiler. The resulting change is small. The bulk of the new code lives in a self-contained package and accounts for approximately 2\,200 new lines, of which about 600 are for the solver (\S\ref{sec:solver}). Changes to existing compiler code are limited to about 300 net lines. The main integration points are:

\begin{itemize}
  \item \emph{Type representation.} Dotty provides a mechanism for defining custom types via type annotations, supporting arbitrary internal representations. We leverage this to attach predicates to types. These predicates are represented using a lightweight expression IR that captures only the subset of Scala expressions valid as predicates (\S\ref{sec:predicates}). Crucially, the leaf nodes of this IR are Scala's existing path-dependent types, so the compiler's existing substitution and dependent-type machinery is reused automatically.
  \item \emph{Parsing.} The parser is extended with the two grammar productions presented in \S\ref{sec:syntax}.
  \item \emph{Subtyping.} A single case is added to the type comparer, corresponding to \textsc{S-Refine} (Figure~\ref{fig:subtyping}): when the supertype is $\{x : A \mid p\}$, the comparer checks $\mathord{<:}\, A$ and delegates predicate entailment to the solver (\S\ref{sec:solver}).
  \item \emph{Adaptation.} Dotty's bidirectional type checker has a single point where a synthesized type is reconciled with an expected type. This point gains one new case: when the expected type is a refinement type, it attempts selfification (\S\ref{sec:inference}), wrapping the expression in $\{x : T \mid x \mathbin{\texttt{==}} e\}$.
  \item \emph{Branch and case facts.} The typer records branch conditions (\lstinline!if!) and pattern-match outcomes in the context, and propagates them to the solver. Unlike the formalization, which also stores equality facts for \lstinline!val! bindings in the typing context (\textsc{T-Let} in Figure~\ref{fig:typing}), the implementation does not need to: the compiler already links each symbol to its definition tree, so the solver can follow references in a predicate and recover definitions on demand, gathering only the facts relevant to the predicate being checked.
  \item \emph{Argument hoisting.} When a method signature contains refinement types that refer to other parameters, the corresponding arguments are lifted to local \lstinline!val! bindings~(\S\ref{sec:anf}). This is needed so that run-time checks can refer to the argument without re-evaluating it.
  \item \emph{Recursive signatures.} When a return type's predicate refers to the method itself, the compiler pre-installs a provisional type without the qualifier to break the cyclic reference.
  \item \emph{Run-time checks.} The type-test lowering phase compiles pattern matches against $\{x : A \mid p\}$ into a type test against~$A$ followed by a run-time evaluation of~$p$~\cite{quentin,valentin}.
\end{itemize}

\subsection{Hoisting and Skolems}\label{sec:anf}

In the formalization (\S\ref{sec:formalization}), \textsc{T-App} (Figure~\ref{fig:typing}) requires the argument to be a variable~$y$, so that the return type $B[x \mapsto y]$ is a simple substitution of one variable for another. In practice, Scala allows arbitrary expressions as arguments, so the implementation must recover this structure on demand through two mechanisms.

\paragraph{Argument hoisting.}
Consider a method \lstinline!def f(x: Int, y: {v: Int with v > x}): Unit!. When called as \lstinline!f(impure(), 3.runtimeChecked)!, the run-time check for the second argument must evaluate \lstinline!impure() > 3!---but \lstinline!impure()! may have side effects and must not be evaluated twice. The compiler detects that \lstinline!x! is referenced from a qualifier in another parameter's type and lifts the argument to a local binding. This transformation is triggered whenever a parameter is referenced from a qualifier in another parameter's type or in the return type.

\paragraph{Skolems as on-the-fly ANF}
How to apply a dependent method \lstinline!f: (x: A) => B! to an arbitrary expression~\lstinline!e!? We cannot blindly substitute, because the return type $B[x \mapsto e]$ may not be defined if \lstinline!e! is outside of the predicate language fragment. This is a well-known problem in refinement type systems. Borkowski et al.~\cite{borkowski2022} solve it with term-level existential types, giving \lstinline!f(e)! the type $\exists x.\, A \land B[x]$ (following Knowles and Flanagan~\cite{flanagan2006}); Hamza et al.~\cite{hamza2019} use a ``let-in-type'' construct $\mathsf{let}\; x = e\;\mathsf{in}\; B$.

Our formalization takes a simpler route: \textsc{T-App} requires the argument to be a variable, which is always valid in a predicate. In the implementation, we achieve the same effect using \emph{skolems}: special terms in predicates that refer to specific but unknown values. Each non-variable argument is assigned a unique skolem \lstinline!?k!. The return type of \lstinline!f(e)! is formed by substituting \lstinline!?k! for \lstinline!x! in \lstinline!B!. If \lstinline!e! happens to live in the supported fragment, an equality \lstinline!?k == e! is added to the equality context.

For example, assume a case class \lstinline!Range(from: Int, until: Int)! whose constructor returns the selfified type \lstinline!{r: Range | r == Range(from, until)}!. When \lstinline!Range(0, 10)! appears as a receiver (say, in \lstinline!Range(0, 10).foreach(...)!), it is assigned a fresh skolem \lstinline!?1!, and \lstinline!foreach! yields elements of type \lstinline!{x: Int | ?1.from <= x && x < ?1.until}!. The solver then gets \lstinline!?1 == Range(0, 10)! as an assumption, allowing it to unfold \lstinline!?1.from! to \lstinline!0! and \lstinline!?1.until! to \lstinline!10!, recovering the bounds needed to verify body expressions like \lstinline!a(i)!.

In effect, the skolem plays the role of the let-bound variable in the formalization: it names the argument value, and the equality assumption carries the necessary fact. Like let bindings in the formalization, skolems are removed from result types via the \lstinline!avoid! operation (\S\ref{sec:typing}) when they would escape their scope.

\subsection{Solver}\label{sec:solver}

Rather than depending on an external SMT solver---which would add platform-specific binaries and a large unowned dependency to the compiler---we built a self-contained solver ($\approx$600 lines) based on e-graphs, lightweight enough to run on every keystroke in an IDE.

The solver decides predicate entailment $P_1 \Rightarrow P_2$, with top-level case splitting for disjunctive assumptions. It collects all relevant facts---qualifier predicates, \lstinline!val! equalities, and branch conditions---inserts them into an e-graph merged with \lstinline!true!, and checks whether $P_2$ becomes equivalent to \lstinline!true!.

\paragraph{Acyclic e-graph.}
Our e-graph implementation is similar to the \emph{acyclic e-graphs} used in the Cranelift compiler~\cite{cranelift} and described by Zucker~\cite{zucker2024}: each equivalence class eagerly collapses to a single canonical representative, rather than maintaining the set of all equivalent terms as in standard e-graphs~\cite{willsey2021}. When two nodes are merged, an \emph{order} function picks the representative (preferring constants, then operations, then constructor applications, then lambdas), and all parent nodes are re-canonicalized. This may trigger further merges---for example, merging $x \equiv 3$ causes $x + y$ to re-canonicalize to $3 + y$, which may match an existing node and cascade into another merge. This repair loop is the core of \emph{congruence closure}~\cite{nelson1980}.

\paragraph{Normalization.}
During canonicalization, the e-graph applies domain-specific rewrites:
\begin{itemize}
  \item \emph{Arithmetic.} Integer sums and products are flattened using associativity, like-terms are grouped and constant-folded ($2x + 3x \to 5x$). Subtraction is rewritten to addition of a negated product ($a - b \to a + (-1) \cdot b$). Distribution is not applied, to avoid exponential blowup.
  \item \emph{Boolean logic.} Conjunction and disjunction short-circuit when one operand is a constant. Double negation is eliminated. When a conjunction is merged with \lstinline!true!, each conjunct is individually merged with \lstinline!true!; dually for disjunction and \lstinline!false!. When an equality $a \mathbin{\texttt{==}} b$ is merged with \lstinline!true!, the nodes $a$ and $b$ are themselves merged.
  \item \emph{Beta reduction.} Applying a lambda to arguments substitutes and re-canonicalizes.
  \item \emph{Constructor--field reduction.} Selecting a field from a constructor application returns the corresponding argument ($\mathsf{Range}(0, 10).\mathtt{from} \to 0$). This is the mechanism that unfolds skolems in the Range example of \S\ref{sec:anf}.
  \item \emph{Transitivity of $<$.} When $a < b$ becomes true, the solver derives $a < c$ for all known $c$ with $b < c$, and symmetrically for lower bounds.
\end{itemize}

\paragraph{Binding.}
E-graph nodes may contain lambdas (for predicates passed to higher-order functions). We represent bound variables with de~Bruijn indices. When checking $P_1 \Rightarrow P_2$, both predicates are opened by substituting their outermost bound variable with a fresh free variable, akin to the opening operation in locally-nameless representations. This is sound because bound variables are never merged with any other node---they remain fully abstract, only equal to themselves. Only the opened free variable participates in merges. This approach avoids the complexity of slotted e-graphs~\cite{schneider2024}, which parameterize e-classes by their free variables to support full equality saturation with binders---a generality we do not need.

\paragraph{Trade-offs.}
This design is strictly more powerful than substitution alone: congruence closure propagates equalities through containing expressions, and cascading repair discovers multi-step simplifications that no single substitution pass would find. It is intentionally less complete than full equality saturation~\cite{willsey2021}: the \emph{order} function picks one representative per class, so if two equivalent forms of a subexpression would trigger different simplifications, only one is explored. We believe this is an acceptable trade-off for a type checker, where predictable performance matters more than completeness.

\subsection{Evaluation}\label{sec:evaluation}

We evaluate our implementation along two axes: compilation time overhead and expressiveness compared to two existing systems for Scala---Schmid and Kun\v{c}ak's previous prototype~\cite{schmid2016} and the Stainless verifier~\cite{hamza2019}. Stainless is not a refinement type system; it checks pre- and postconditions. It is, however, the de facto tool for verifying Scala programs, even for lightweight properties that refinement types could cover. This makes it a natural point of comparison for our system, on the programs that both systems can verify.

\paragraph{Compilation overhead.}
We measure single-shot compilation time on a suite of small benchmark programs, comparing each system against its own baseline: the same programs without refinement annotations, compiled with checking disabled. Where possible, we port the same program to Schmid's system and to Stainless. Baselines are per-system because the systems build on different compiler versions and the ported programs differ slightly in code shape. Benchmarks run under JMH; each benchmark is compiled with 150 single-shot warm-up iterations followed by 20 measurement iterations, and the whole experiment is repeated in 10 independent runs. Each run is reduced to the mean of its measurement iterations; Table~\ref{tab:bench-table} reports the mean of the per-run means with 95\% confidence intervals (Student's $t$), and the overhead~$\Delta$ as the difference between checked and baseline times.

On our benchmarks, our system (labelled ``First-class'' in the table) adds modest overhead: 0--12\% of the baseline compilation time. This is consistently less than the two other systems: on the benchmarks they support, Schmid's system adds 20--38\% and Stainless at least 56\%.

To gauge overhead on a realistic codebase, we applied our system to \emph{Fansi}~\cite{fansi}, a 1\,000-line terminal-color library. The only change was replacing three run-time \lstinline!require! checks in a color-index function with a refinement type alias:
\begin{lstlisting}[language=Scala]
type ChannelValue = {v: Int with 0 <= v && v < 256}
def trueIndex(r: ChannelValue, g: ChannelValue,  b: ChannelValue)
\end{lstlisting}
All call sites already guard the arguments with the same bounds, so the refinement is verified statically with no other code changes. The compilation time difference is within measurement noise, confirming that the overhead is proportional to the number of refinement checks, not to the size of the program. This experiment also illustrates a key advantage of first-class integration: our system compiles unmodified Scala code without restriction, so refinement types can be introduced incrementally into an existing codebase. By contrast, Stainless is limited to a specific fragment of Scala and fails on code that falls outside its supported subset.

\paragraph{Impact on code that does not use the feature.}
Because refinement checking is integrated into the compiler rather than layered on top, a natural concern is whether it affects existing code. To assess this, we run the full Dotty continuous-integration suite with the feature enabled. It passes unchanged: this includes compiling the Scala compiler itself (${\sim}200\,000$ lines of code), its ${\sim}10\,000$ compilation tests (${\sim}300\,000$ lines), and a community build of 50 open-source projects (${\sim}700\,000$ lines), indicating that the extension introduces no backward incompatibility. On the official Scala compiler benchmark suite,\footnote{\url{https://lampepfl.github.io/scala3-benchmarks/}} enabling the feature slows compilation by at most 4\% with the current prototype, on both the 11 ``Big'' benchmarks, which compile real-world libraries totalling ${\sim}70\,000$ lines, and the 20 ``Small'' benchmarks. In short, when the feature is unused, it changes no behavior and adds only a small compilation-time overhead.

\begin{table*}[t]
\centering
\captionsetup{font=small}
\caption{Compilation time benchmarks (ms/op, single-shot). Each run (JMH fork) is reduced to its mean; scores are the mean of the per-run means $\pm$ the 95\% confidence interval across runs (Student's $t$). Each platform shows the unchecked baseline time and the absolute overhead of checking; the checked time is $\text{base} + \Delta$. Since the baseline and checked samples are independent, the overhead interval is $\sqrt{\smash[b]{e_b^2 + e_c^2}}$, where $e_b$ and $e_c$ are the baseline and checked intervals.}
\label{tab:bench-table}
\footnotesize
\begin{tabular}{l rrr rrr rrr}
\toprule
 & \multicolumn{3}{c}{First-class} & \multicolumn{3}{c}{Schmid} & \multicolumn{3}{c}{Stainless} \\
\cmidrule(lr){2-4} \cmidrule(lr){5-7} \cmidrule(lr){8-10} 
Benchmark & LoC & Base (ms) & $\Delta$ (ms) & LoC & Base (ms) & $\Delta$ (ms) & LoC & Base (ms) & $\Delta$ (ms) \\
\midrule
\texttt{arrfold} & $20$ & $94 \pm 1$ & $3 \pm 2$ & $20$ & $88 \pm 2$ & $30 \pm 2$ & --- & --- & --- \\
\texttt{collect} & $15$ & $103 \pm 2$ & $4 \pm 3$ & --- & --- & --- & --- & --- & --- \\
\texttt{fansi} & $530$ & $187 \pm 3$ & $-1 \pm 3$ & --- & --- & --- & --- & --- & --- \\
\texttt{fibMemo} & $15$ & $104 \pm 1$ & $7 \pm 2$ & --- & --- & --- & --- & --- & --- \\
\texttt{hof1} & $5$ & $75 \pm 2$ & $3 \pm 2$ & $5$ & $77 \pm 1$ & $18 \pm 1$ & --- & --- & --- \\
\texttt{hof2} & $8$ & $77 \pm 1$ & $2 \pm 1$ & $8$ & $77 \pm 1$ & $20 \pm 1$ & --- & --- & --- \\
\texttt{intarray} & $11$ & $94 \pm 1$ & $3 \pm 1$ & $11$ & $75 \pm 1$ & $19 \pm 2$ & $14$ & $68 \pm 1$ & $45 \pm 2$ \\
\texttt{list1} & $5$ & $95 \pm 1$ & $3 \pm 2$ & $5$ & $83 \pm 2$ & $17 \pm 2$ & --- & --- & --- \\
\texttt{list2} & $5$ & $95 \pm 1$ & $2 \pm 2$ & $5$ & $83 \pm 2$ & $17 \pm 3$ & --- & --- & --- \\
\texttt{matrix} & $51$ & $101 \pm 2$ & $11 \pm 2$ & --- & --- & --- & --- & --- & --- \\
\texttt{matrixDims} & $13$ & $94 \pm 2$ & $4 \pm 2$ & $13$ & $75 \pm 2$ & $31 \pm 2$ & $21$ & $66 \pm 1$ & $93 \pm 1$ \\
\texttt{max} & $4$ & $70 \pm 1$ & $5 \pm 2$ & $4$ & $74 \pm 1$ & $25 \pm 2$ & $6$ & $49 \pm 1$ & $56 \pm 2$ \\
\texttt{maxGen} & $11$ & $103 \pm 1$ & $2 \pm 2$ & --- & --- & --- & --- & --- & --- \\
\texttt{maxGenBig} & $15$ & $109 \pm 2$ & $14 \pm 2$ & --- & --- & --- & --- & --- & --- \\
\texttt{mergeSort} & $29$ & $95 \pm 1$ & $11 \pm 2$ & --- & --- & --- & $53$ & $62 \pm 2$ & $648 \pm 4$ \\
\texttt{postuple} & $4$ & $71 \pm 2$ & $6 \pm 2$ & $4$ & $73 \pm 1$ & $18 \pm 2$ & $9$ & $68 \pm 1$ & $36 \pm 1$ \\
\texttt{range} & $19$ & $99 \pm 1$ & $4 \pm 2$ & --- & --- & --- & --- & --- & --- \\
\texttt{rational} & $6$ & $71 \pm 1$ & $5 \pm 2$ & $6$ & $75 \pm 1$ & $26 \pm 1$ & $10$ & $67 \pm 1$ & $47 \pm 1$ \\
\texttt{sqrt} & $9$ & $74 \pm 1$ & $7 \pm 2$ & $9$ & $77 \pm 1$ & $25 \pm 1$ & $14$ & $49 \pm 1$ & $66 \pm 2$ \\
\texttt{sumnat} & $10$ & $69 \pm 1$ & $6 \pm 1$ & $10$ & $74 \pm 1$ & $21 \pm 2$ & --- & --- & --- \\
\texttt{vec} & $12$ & $74 \pm 2$ & $6 \pm 2$ & --- & --- & --- & $20$ & $56 \pm 1$ & $136 \pm 2$ \\
\bottomrule
\end{tabular}
\end{table*}

\paragraph{Expressiveness.}
The three systems differ in which programs they can express. Schmid's qualifier inference operates in a separate phase from Scala's type checker, so refinements do not participate in type-argument inference: generic data structures can be instantiated with qualified types only by writing the type arguments explicitly, and inferred instantiations such as the one in Figure~\ref{fig:example1} are out of reach. Stainless supports richer specifications through \lstinline!require! and \lstinline!ensuring! clauses with a full SMT backend, but refinements are not types: they do not compose with bounded polymorphism, type inference, or pattern matching. Our system integrates with all of these by construction, since refinement types are ordinary Scala types.

On the other hand, both Schmid's system and Stainless have substantially more powerful solvers. Both use SMT-backed solvers that are complete for linear integer arithmetic, algebraic data types, equality, and first-order logic---theories for which our lightweight e-graph solver offers no decision procedure. As a consequence, several of Schmid's benchmarks that require linear arithmetic reasoning (e.g., \texttt{sumnat}, \texttt{arrfold}) cannot be verified by our solver, and Stainless can verify functional correctness properties well beyond what refinement types alone can express.

\section{Related Work}\label{sec:related-work}

\paragraph{Mechanized refinement type systems.}
Hamza et~al.~\cite{hamza2019} mechanize System~FR, a dependently-typed calculus with refinement types, polymorphism, recursive types, and dependent pairs, serving as a foundation for the Stainless verifier. Their system is the closest to ours in ambition and was a major inspiration, particularly for the positivity-based treatment of recursive types. Key differences are: (1)~System~FR enforces \emph{total correctness}, requiring termination proofs for all recursive functions, while our partial-correctness approach avoids this burden; (2)~System~FR uses dependent pairs as existentials to handle variables going out of scope, while we use avoidance; and (3)~System~FR uses index-guarded recursive types $\textsf{Rec}(n)(\alpha \Rightarrow \tau)$, while we use equi-recursive types with fold/unfold subtyping.
Borkowski et~al.~\cite{borkowski2022} mechanize a refinement type system extending System~F in Liquid Haskell. They restrict refinements to base types (no nested refinements), use well-formedness conditions and existential types rather than avoidance, and elegantly axiomatize the implication judgment as eleven abstract properties that an SMT oracle must satisfy for soundness, without committing to a specific solver. We instead leave implication fully semantic (\S\ref{sec:soundness}).
Sun et~al.~\cite{sun2024} mechanize RFJ, a refinement type system for a Featherweight Java-like language, in Coq. They provide both type soundness and logical soundness (characterizing which formulas are valid). Their system lacks parametric polymorphism but handles selfification and class-based subtyping in an object-oriented setting.

\paragraph{Definitional interpreters and semantic soundness.}
Our mechanization builds on Amin and Rompf's~\cite{amin2017} technique of proving type safety via a fuel-bounded definitional interpreter. One of their key insights is that big-step semantics sidesteps the ``bad bounds'' problem that plagues small-step proofs for languages with type members. However, their development proves syntactic type safety (progress and preservation) rather than defining semantic types.
Owens et~al.~\cite{owens2016} introduced functional big-step semantics, and Paraskevopoulou et~al.~\cite{paraskevopoulou2021} use a similar interpreter-based approach with step-indexed logical relations for multi-pass compiler verification. In both systems, fuel is consumed at every evaluation step, whereas ours bounds evaluation depth (\S\ref{para:fuel}).
Semantic typing, interpreting types as predicates or relations on values, has a long history in logical relations~\cite{timany2024}. Giarrusso et~al.~\cite{giarrusso2020} develop a semantic model for DOT, with subtyping and recursive types using step-indexed logical relations and a ``later'' modality in the type syntax. We avoid step-indexing entirely by requiring strict positivity for recursive type binders (following Hamza et~al.~\cite{hamza2019}), enabling a direct fixpoint construction via intersected approximations (\S\ref{sec:soundness}).

\paragraph{Refinement types in other languages.}
We reviewed the history of refinement types in the introduction. Flanagan's hybrid type checking~\cite{flanagan2006} introduced the idea of automatically inserting run-time checks when the static system cannot discharge a refinement obligation. In our system, run-time checks are not inserted automatically: the programmer explicitly requests them via \lstinline!.runtimeChecked! or pattern matching (\S\ref{sec:design}), which the type system treats as an ordinary conditional branch, so no special rules are needed for the static/dynamic boundary. Moreover, our run-time checks are restricted to first-order predicates, which avoids the blame assignment problem that arises when monitoring higher-order contracts~\cite{keil2015}: since functions cannot be checked eagerly, a violation detected at a later call site must be attributed to the party at fault.

Liquid Haskell~\cite{vazou2014, vazou2017} brings refinement types to Haskell as a separate plugin, with refinement inference running apart from GHC's own type checking, whereas our refinements are typed and inferred by the compiler itself. Termination also plays a different role: Liquid Haskell by default requires every function to be proven terminating, while we need termination only in the solver, to justify certain normalization and unfolding rules (\S\ref{sec:termination}). Finally, Liquid Haskell discharges obligations with an SMT solver over decidable theories, while our solver (\S\ref{sec:solver}) combines e-graphs with normalization rules: it forgoes arithmetic decision procedures, but natively performs beta reduction and limited higher-order reasoning over predicate lambdas. The recent PLEX extension~\cite{ferrarini2026} adds normalization---eta-expansion, beta-reduction, and dependent pattern matching---to Liquid Haskell's Proof by Logical Evaluation layer, precisely to recover higher-order reasoning over a first-order SMT solver.

In the systems programming space, refinement types have been combined with mutation, where the central difficulty is deciding when established predicates must be invalidated: a refinement of a mutable location can be treated as an \emph{invariant} that every write must preserve, or as a flow-sensitive fact that a write updates or invalidates. Under aliasing, knowing which writes affect which facts is itself hard.
Refinement type systems for Rust lean on the ownership discipline: Flux~\cite{lehmann2023} permits \emph{strong updates}, changing a location's refinement, through exclusively owned pointers, but only invariant-preserving \emph{weak updates} through mutable borrows, while Thrust~\cite{thrust2025} and RefinedRust~\cite{gaher2024} type a borrow as its current value paired with a \emph{prophecy variable} for its final value.
RefinedC~\cite{sammler2021} and RefinedRust come with foundational soundness proofs in the Iris separation logic, and Verus~\cite{lattuada2023} provides SMT-based verification for Rust using linear ghost types; these systems target memory safety and low-level invariants. Generic refinement types~\cite{lehmann2025} add refinement polymorphism to Flux.
The separation checking sketched in \S\ref{sec:purity} would give Scala a similar handle on invalidation: like ownership, capabilities~\cite{xu2025} with separation checking~\cite{xu2026} make writes explicit and un-aliased, identifying the facts a write can invalidate. For TypeScript, Refined TypeScript~\cite{vekris2016} sidesteps invalidation by allowing refinements to mention object state only through immutable fields, and recovers flow-sensitive reasoning for local variables via an SSA translation.

The Checker Framework~\cite{dietl2011} brings user-defined \emph{pluggable} type qualifiers to Java: like refinement types, qualifiers attach additional information to types, and like the separate-phase systems above, the checkers run as compiler plugins over Java's annotation syntax, outside the language's own type system. Qualifiers are drawn from a fixed vocabulary and checked by dataflow analysis rather than a solver, a trade-off that scales to large codebases and many practical properties, such as verifying array indexing~\cite{kellogg2018}, at the cost of expressiveness. Lanzinger et~al.~\cite{lanzinger2021} blur the line between the two approaches by combining pluggable type qualifiers with deductive verification, discharging obligations that the type checker cannot prove with a program verifier.

\section{Conclusion}\label{sec:conclusion}

We have shown that refinement types can be integrated as first-class citizens into the type system of a mainstream programming language. By extending the Scala~3 compiler directly, refinement types interact naturally with subtyping, bounded polymorphism, type inference, and pattern matching, avoiding the two-mental-models problem of separate-phase approaches.

On the theoretical side, we provide the first mechanized soundness proof that combines refinement types with union and intersection types, bounded polymorphism with lower and upper bounds, and positive equi-recursive types. The proof uses a fuel-bounded definitional interpreter with semantic typing in Rocq, and establishes partial-correctness type safety for a core calculus that captures key features of Scala's type system. On the practical side, the prototype adds refinement types to an industrial-strength compiler with modest changes ($\approx$2\,500 lines), a self-contained e-graph solver requiring no external dependencies, and compilation overhead that remains low.

\FloatBarrier

\clearpage

\section*{Data-Availability Statement}

The accompanying artifact~\cite{bovel_2026_21737784} contains:

\begin{itemize}
  \item \textbf{Mechanization.} The Rocq formalization of the type system, with its soundness proof.
  \item \textbf{Implementation.} The source of our Dotty extension, also published on Maven and usable from any SBT project with:
\begin{lstlisting}[language=Scala]
scalaOrganization := "ch.epfl.lara"
scalaVersion := "3.10.0-RC1-bin-20260720-5f1bae1-NIGHTLY"
scalacOptions ++= Seq("-language:experimental.qualifiedTypes")
\end{lstlisting}
  \item \textbf{Benchmarks.} The JMH benchmark suites used for the evaluation, compiling similar programs with our Dotty fork, Schmid's prototype, and Stainless, together with scripts to run them and collect results.
\end{itemize}

The same material, with its continued development, is available in the project's public repository: \url{https://github.com/mbovel/first-class-refinement-types-in-scala}.

\subsection*{AI-Usage Statement}

AI tools were used in the preparation of this paper and its artifacts, primarily Claude Code (Opus 4.5 and 4.6, Fable 5). Theorem and lemma statements were written by the authors, while many proof bodies were generated with AI assistance. The core implementation predates widespread agentic AI tools, but recent bug fixes and refactorings were AI-assisted (AI-assisted commits are explicitly marked in the git history with \texttt{Co-authored-by: Claude *}). GitHub Copilot was also used for code autocompletion, and Google and ChatGPT for general research questions.

\begin{acks}

We thank the LAMP and LARA teams at EPFL for insightful discussions and the prior work that laid the foundation for this paper, and in particular Sébastien Doeraene and Guillaume Martres for their valuable feedback on the Dotty implementation.

\end{acks}

\bibliographystyle{ACM-Reference-Format}
\bibliography{references}

\makeatletter
\par\bigskip\noindent{\small\normalfont\@received\par}%
\gdef\@received{}%
\makeatother

\clearpage

\appendix


\section{More examples}

\begin{lstlisting}[language=Scala]
// Example 3: Matrix operations with checked dimensions
case class Matrix(width: Pos, height: Pos):
  private val size = width * height
  private val data = CheckedArray(size.runtimeChecked)
  def index(i: Pos with i < height,
            j: Pos with j < width): {res: Pos with res < data.size} =
    (i * width + j).runtimeChecked
  def apply(i: Pos with i < height, j: Pos with j < width): Double =
    data(index(i, j))
  def update(i: Pos with i < height, j: Pos with j < width, x: Double): Unit =
    data(index(i, j)) = x
  def transpose(): {v: Matrix with v.width == height && v.height == width} = ???
  def mul(that: Matrix with that.height == width):
    {v: Matrix with v.width == that.width && v.height == height}
  =
    val res = Matrix(that.width, height)
    forRange(0, height): i =>
      forRange(0, that.width): j =>
        var sum = 0.0
        forRange(0, width): k =>
          sum = sum + this(i, k) * that(k, j)
        res(i, j) = sum
    res
def matricesExample =
  val m1 = Matrix(2, 3)
  val m2 = Matrix(3, 2)
  m1.mul(m2)
  val m1T = m1.transpose()
  m1T.mul(m1)

// Example 4: Memoized Fibonacci with a dependent map
case class DepMap[K, V](p: (K, V) => Boolean):
  def put(n: K, v: V with p(n, v)): Unit = ???
  def get(n: K): Option[{res: V with p(n, res)}] = ???
def fib(n: Int): {r: Int with r == (if n <= 1 then 1 else fib(n - 1) + fib(n - 2))} =
  if n <= 1 then 1 else fib(n - 1) + fib(n - 2)
val cache: (DepMap[Int, Int] with (cache == DepMap[Int, Int]((k, v) => v == fib(k)))) =
  DepMap[Int, Int]((k, v) => v == fib(k))
def fibMemo(n: Int): {res: Int with res == fib(n)} =
  cache.get(n) match
    case Some(res) => res
    case None =>
      val res: (Int with (res == fib(n))) =
        if n <= 1 then 1 else fibMemo(n - 1) + fibMemo(n - 2)
      cache.put(n, res)
      res
\end{lstlisting}

\end{document}